\newif\iffigs\figsfalse
\figstrue

\documentstyle[12pt]{article}

\iffigs
  \input epsf
\else
\fi

\batchmode
  \newfont{\footscrfont}{rsfs10}
  \newfont{\footbbbfont}{msbm10}
\errorstopmode

\newif\ifscrf\scrftrue
\ifx\footscrfont\nullfont
  \scrffalse
\fi

\newif\ifamsf\amsftrue
\ifx\footbbbfont\nullfont
  \amsffalse
\fi

\def\ppnumber{IASSNS-HEP-93/49}
\def\ppdate{October, 1993}
\def\pplogo{\vbox{\kern-\headheight\kern -1pt
\halign{##&##\hfil\cr&{
\ppnumber}\cr\rule{0pt}{2.5ex}&\ppdate\cr}
}}

\makeatletter
\date{}
\def\dedicatory#1{\def\@date{\normalsize\it#1}}
\def\subjclass#1{\def\@thefnmark{}\@footnotetext{1991
    {\it Mathematics Subject Classification.} #1}}
\def\keywords#1{\def\@thefnmark{}\@footnotetext{
    {\it Key words and phrases.} #1}}

\def\ps@firstpage{\ps@empty \def\@oddhead{\hss\pplogo}%
  \let\@evenhead\@oddhead 
}
\def\maketitle{\par
 \begingroup
 \def\thefootnote{\fnsymbol{footnote}}
 \def\@makefnmark{\hbox
 to 0pt{$^{\@thefnmark}$\hss}}
 \if@twocolumn
 \twocolumn[\@maketitle]
 \else \newpage
 \global\@topnum\z@ \@maketitle \fi\thispagestyle{firstpage}\@thanks
 \endgroup
 \setcounter{footnote}{0}
 \let\maketitle\relax
 \let\@maketitle\relax
 \gdef\@thanks{}\gdef\@author{}\gdef\@title{}\let\thanks\relax}

\def\abstract{\if@twocolumn
\section*{Abstract}
\else \small
\begin{center}
{\bf ABSTRACT}
\end{center}
\quotation
\fi}

\def\thebibliography#1{\section*{References\@mkboth
 {REFERENCES}{REFERENCES}}\small\list
 {\arabic{enumi}.}{\settowidth\labelwidth{[#1]}\leftmargin\labelwidth
 \advance\leftmargin\labelsep
 \usecounter{enumi}}
 \def\newblock{\hskip .11em plus .33em minus .07em}
 \sloppy\clubpenalty4000\widowpenalty4000
 \sfcode`\.=1000\relax}

\newif\iffn\fnfalse

\@ifundefined{reset@font}{\let\reset@font\empty}{} 
\long\def\@footnotetext#1{\insert\footins{\reset@font\footnotesize
    \interlinepenalty\interfootnotelinepenalty
    \splittopskip\footnotesep
    \splitmaxdepth \dp\strutbox \floatingpenalty \@MM
    \hsize\columnwidth \@parboxrestore
   \edef\@currentlabel{\csname p@footnote\endcsname\@thefnmark}\@makefntext
    {\rule{\z@}{\footnotesep}\ignorespaces
      \fntrue#1\fnfalse\strut}}}
\makeatother

\ifamsf
  \newfont{\bigbbbfont}{msbm10 scaled\magstep2}
  \newfont{\bbbfont}{msbm10 scaled\magstep1}  
  \newfont{\smallbbbfont}{msbm8}
  \newfont{\tinybbbfont}{msbm6}
  \newfont{\smallfootbbbfont}{msbm7}
  \newfont{\tinyfootbbbfont}{msbm5}
\fi

\ifscrf
  \newfont{\scrfont}{rsfs10 scaled\magstep1}  
  \newfont{\smallscrfont}{rsfs7}
  \newfont{\tinyscrfont}{rsfs7}
  \newfont{\smallfootscrfont}{rsfs7}
  \newfont{\tinyfootscrfont}{rsfs7}
\fi

\ifamsf
  \newcommand{\Bbb}[1]{\iffn
      \mathchoice{\mbox{\footbbbfont #1}}{\mbox{\footbbbfont #1}}
      {\mbox{\smallfootbbbfont #1}}{\mbox{\tinyfootbbbfont #1}}\else
      \mathchoice{\mbox{\bbbfont #1}}{\mbox{\bbbfont #1}}
      {\mbox{\smallbbbfont #1}}{\mbox{\tinybbbfont #1}}\fi}
\else
  \def\bigbbbfont{\bf}
  \def\Bbb{\bf}
\fi

\ifscrf
  \newcommand{\Scr}[1]{\iffn
    \mathchoice{\mbox{\footscrfont #1}}{\mbox{\footscrfont #1}}
    {\mbox{\smallfootscrfont #1}}{\mbox{\tinyfootscrfont #1}}\else
    \mathchoice{\mbox{\scrfont #1}}{\mbox{\scrfont #1}}
    {\mbox{\smallscrfont #1}}{\mbox{\tinyscrfont #1}}\fi}
\else
  \def\Scr{\cal}
\fi

\newcommand{\text}[1]{\mathchoice{\mbox{\rm #1}}{\mbox{\rm #1}}
    {\mbox{\scriptsize\rm #1}}{\mbox{\tiny\rm #1}}}
\newcommand{\operatorname}[1]{\mathop{\rm #1}\nolimits}

\newcommand{\eqref}[1]{(\ref{#1})}

\newcommand{\C}{{\Bbb C}}
\newcommand{\F}{{\cal F}}

\renewcommand{\P}{{\Bbb P}}

\newcommand{\R}{{\Bbb R}}
\newcommand{\Z}{{\Bbb Z}}

\newcommand{\Area}{\operatorname{Area}}
\newcommand{\Vol}{\operatorname{Vol}}

\newcommand{\tr}{\operatorname{tr}}

\newcommand{\Img}{\operatorname{Im}}

\def\opeq#1{\advance\lineskip#1 \advance\baselineskip#1
	\advance\lineskiplimit#1}
\def\eqalign#1{\null\,\vcenter{\opeq{2.5\jot}\mathsurround=0pt
	\everycr={}\tabskip=0pt
	\halign{\strut\hfil$\displaystyle{##}$&$\displaystyle{{}##}$\hfil
	\crcr#1\crcr}}\,\null}

\def\sm{$\sigma$-model}
\def\smm{\sm\ measure}
\def\CY{Calabi-Yau}
\def\cK{{K}}
\def\cR{{\Scr R}}
\def\cM{{\Scr M}}
\def\cA{{\Scr A}}
\def\cF{{\Scr F}}
\def\cD{{\Scr D}}
\def\SLG{S_{\hbox{\scriptsize LG}}}
\def\cMu{\cM}
\def\cMc{{\hfuzz=100cm\hbox to 0pt{$\;\overline{\phantom{X}}$}\cM}}
\def\mKf{algebraic measure}
\def\ff#1#2{{\textstyle\frac{#1}{#2}}}
\def\F#1#2{{}_{#1}F_{#2}}

\begin{document}
\setcounter{page}0
\title{\LARGE Measuring Small Distances in\\
$N$=2 Sigma Models\\[5mm]
}
\author{\vbox{
\begin{tabular}{c@{\hspace{2cm}}c}
\normalsize Paul S. Aspinwall& \normalsize Brian R. Greene\thanks{On
leave from F. R. Newman Laboratory of Nuclear Studies, Cornell University,
 Ithaca, NY 14853}\\
\normalsize School of Natural Sciences&
\normalsize School of Natural Sciences\\
\normalsize Institute for Advanced Study&
\normalsize Institute for Advanced Study\\
\normalsize Princeton, NJ  08540&
\normalsize Princeton, NJ  08540
\end{tabular}}\\ \null\\[2mm]
\normalsize David R. Morrison\thanks{On leave from
Department of Mathematics, Duke University, Durham, NC 27708} \\
\normalsize School of Mathematics \\
\normalsize Institute for Advanced Study\\
\normalsize Princeton, NJ 08540
}

{\hfuzz=10cm\maketitle}

\renewcommand{\Large}{\large}
\renewcommand{\LARGE}{\large\bf}

\begin{abstract}

We analyze global aspects of the moduli space of K\"ahler forms for $N$=(2,2)
conformal $\sigma$-models.  Using algebraic methods and mirror symmetry we
study extensions of the mathematical notion of length (as specified by a
K\"ahler structure) to conformal field theory and calculate the way in which
lengths change as the moduli fields are varied along distinguished paths in
the moduli space.  We find strong evidence supporting the notion that, in the
robust setting of quantum Calabi-Yau moduli space, string theory restricts the
set of possible K\"ahler forms by enforcing ``minimal length'' scales,
provided that topology change is properly taken into account.  Some lengths,
however, may shrink to zero.  We also compare stringy geometry to classical
general relativity in this context.

\end{abstract}

\vfil\break

\section{Introduction}

Geometrical concepts play a central role in our theoretical
descriptions of the fundamental properties of elementary particles
and the spacetime arena within which they interact. The advent of string
theory has reinforced this reliance on geometrical methods; it has done
so, though, with a fascinating twist. String theory {\it necessitates}
the introduction of particular modifications of standard geometrical
constructions which can drastically modify their properties when
the typical length scales involved approach Planckian values.
Conversely, when all length scales involved are large compared to
the Planck scale, these modified geometrical constructs approach their
classical counterparts. This phenomenon of string deformed classical
geometry is usually referred to as ``quantum geometry'' (although
``stringy geometry'' might be more accurate since our concern will
be exclusively at string tree level).

Recently \cite{AGM:I,AGM:II,W:phase},
a striking property of
quantum geometry was uncovered in the context of string theory compactified on
a \CY\  space.  A classical analysis instructs us to limit our attention
to Riemannian metrics on such a space -- that is, to positive definite
bilinear forms mapping $T_X \times T_X$ to $\R^+$. As \CY\ spaces are
K\"ahler, this condition can be rephrased as the statement that the
K\"ahler form on the \CY\ space $X$ lies in a subset of $H^2(X,\R)$
known as the K\"ahler cone. However,
an analysis  based on string theory reveals
a different story. Namely,  it was shown in \cite{AGM:I,AGM:II,W:phase}
that the physics of string theory continues to make perfect sense
even if we allow the ``K\"ahler form'' to take values outside of the
K\"ahler cone. This was shown, for example, to give rise to
physical processes resulting in a change in the topology of the
\CY\ target space -- processes which classical reasoning would
forbid.

Another striking aspect of quantum geometry is the apparent existence
of a minimum length set by the string scale $\alpha^\prime$.
The evidence
for this has come from a variety of studies. First, it has long been
known \cite{YY:} that string theory compactified on a circle of radius
$R$ is physically identical to the theory compactified on a circle of
radius $\alpha^\prime/R$.
The full set of physically distinct
possibilities with this topology is therefore parameterized by
radii $R$ varying from $(\alpha^\prime)^{1/2}$ to $\infty$;
in this sense
$(\alpha^\prime)^{1/2}$ is a minimum length in this setting. Additional
evidence for a minimum length was given in \cite{CDGP:} in which the
one  dimensional space of K\"ahler forms
on the quintic threefold was studied by means of mirror symmetry.
Those authors found that
physically distinct theories are again characterized by K\"ahler forms
which attain
a minimal nonzero volume. From another point of view, the work of
\cite{GM:} showed that there appears to be a smallest length scale
that can be probed via high energy scattering with an extended object
such as a string. Roughly, unlike what happens in the  point particle
case, increasing the energy of the string probe beyond a critical value
results in an increase in the size of the probe itself and hence a
{\it decrease\/} in the length scale of sensitivity.

At first sight, the observations in the last two paragraphs might seem
to be at odds. On the one hand, we have mentioned work which establishes
that string theory {\it relaxes\/} constraints on the \CY\ metric and hence
makes all of $H^2(X,\R)$ available for consistent physical models.
On the other hand, we have referred to work which establishes that string
theory {\it restricts\/} the physically realizable metrics to a subset
of those which are classically allowed. One of the main purposes of the
present paper is to study this issue in some detail and show
the harmonious coexistence of these apparently divergent statements.

As part of the analysis in the sequel is  somewhat technical, it
is worthwhile for us to briefly summarize our results here. To do so,
let us first recall that in \cite{AGM:I,AGM:II,W:phase}
it was shown that string
theory instructs us to pass from
the moduli space of K\"ahler forms on a  single \CY\ space to the
{\it enlarged\/} K\"ahler moduli space. The latter is a space
which comes equipped with a decomposition into cells, each of which
corresponds to a different ``phase'' of the $N = 2$ superconformal theory
(see figure \ref{fig:fo}).
 From a mathematical point of view, one might say the walls between
these cells correspond to K\"ahler forms which degenerate in some manner.
Some of these phases are interpretable in terms of strings
on (birationally equivalent\footnote{We remind the reader that two spaces
are birationally equivalent if upon removing suitable subsets of codimension
one from each they become isomorphic.}
 but possibly topologically distinct) smooth
\CY\ manifolds,  some other phases correspond to strings on singular
(orbifold) \CY\ spaces and yet other phases include Landau-Ginzburg
theories and exotic hybrid combinations.
More precisely, each cell contains a neighbourhood of a distinguished
``limit''  point
(marked with a dot in figure \ref{fig:fo}) around which some kind of
perturbation theory
converges and the above identifications can unambiguously be made.
(For the Calabi-Yau phases, these are known as ``large radius limit'' points.)
The region of convergence is shown by a dotted line in figure
\ref{fig:fo}.
 A generic path in this
enlarged K\"ahler moduli space corresponds to a family of well defined
conformal theories and hence there is no obstruction to passing from
one cell into another. This gives rise to the topology changing transitions
mentioned earlier. Under mirror symmetry, this enlarged K\"ahler moduli
space corresponds to the complex structure moduli space of the mirror.
As discussed in \cite{AGM:II}, the
badly behaved conformal field theories
form a subspace of {\it complex\/} codimension one (as
opposed to the {\it real\/} codimension  one walls in the K\"ahler space)
in an appropriate
compactification of the moduli space, which under mirror symmetry corresponds
to the ``discriminant locus'' of the complex structure moduli space.
As this locus has real codimension
two, a generic path in that
moduli space avoids it.
This is, in fact, how we established that the same must be true
for a generic path in the enlarged K\"ahler moduli space of the original
\CY\ manifold.

\iffigs
\begin{figure}
  \centerline{\epsfxsize=7cm\epsfbox{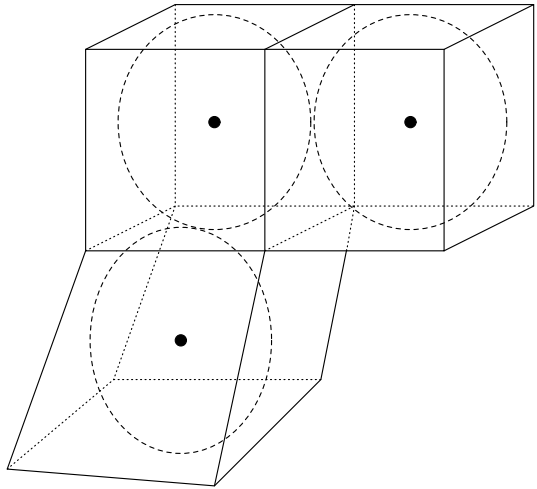}}
  \caption{The cell decomposition of part of the moduli space.}
  \label{fig:fo}
\end{figure}
\fi

Taking this picture at face value,  it appears that some
points in figure 1
correspond to K\"ahler forms with zero or even negative  volumes
(since we pass outside of a single classical K\"ahler cone).
One superficial way of treating this is simply to assert that a geometrical
interpretation can only be given to a subset of points in the moduli space ---
those points with a large positive volume according to some birational
model of the space. Although that point of view avoids the obvious difficulties
about negative volumes, our goal in this paper is to
probe the issue more deeply and determine to what extent we can give
a consistent geometrical interpretation  (and hence
assign a positive volume) to all points in the
enlarged moduli space. A crucial ingredient in such a study is the
precise definition of ``volume'' or ``size'' in the conformal field
theory context. As the size of a space is an inherently classical
mathematical notion, there is no unique way of extending its definition
to quantum geometry. There are, however, a couple of compelling
extensions which are both natural from the point of view of conformal
field theory and which reduce to the standard notion of size in
the appropriate large radius limits.
One of these extensions relies upon mirror symmetry to rewrite the moduli
space of
K\"ahler forms on $X$ as the moduli space of complex structures of
another space $Y$. The coordinates in this moduli space
(which are coupling constants in the action of the associated
conformal field theory)
are then used to
represent the size of $X$ in the simplest possible way
(as we will discuss). This turns out to be
equivalent to measuring ``size'' by using
the classical K\"ahler form on the nonconformal linear
 \sm\ which was studied in
\cite{W:phase}.

The second version of size is derived directly from properties of the
conformal nonlinear
\sm.  This definition can
be obtained by requiring
that it not only approach the notion of size based upon
the classical K\"ahler form
at the large radius limit, but also that it exactly match that notion
 in a certain
neighbourhood of this limit. (This neighbourhood will be the
region in which we can, at least in principle, calculate the conformal
\sm\ correlation functions and thus use the \sm\ as the link between
points in the moduli space and the geometry of $X$.) The measurement of
size is then
analytically continued in a natural way beyond this neighbourhood of
the large radius limit point.
In practice we will analyze this second definition of size by means of
the first definition in the preceding paragraph, and of a function which
relates the two sizes.
This function can be expressed in terms of solutions to
a set of differential equations --- the Picard-Fuchs equations.

The ``sizes'' on which we focus in this paper will be described by
specifying an area\footnote{More precisely, we specify a ``complexified
area'' whose imaginary part is the ordinary area.}
for every holomorphically embedded Riemann surface $C$ in $X$,
or more generally, for every $2$-cycle $C$ on $X$.  We will refer to
such a specification of areas as a {\it measure\/} on $X$, and we will
give precise definitions of the ``\mKf'' (the first notion
of size) and the ``\smm'' (the second notion of size) later in the
paper.

The areas that we specify only depend on the homology class of the Riemann
surface.  If we choose $2$-cycles $C_i$ forming an integral basis of
$H_2(X)$, and let $e^i$ be the dual basis of $H^2(X)$ then associating
something like a
complexified
K\"ahler form
$B + iJ = \sum t_i e_i$
to each conformal field theory in the moduli space, we see
\begin{equation}
  \Area(C_i) = \Img\int_{C_i} (B+iJ) = \Img(t_i).
\end{equation}
One should note that although our moduli space usually contains
theories corresponding to many
topologically distinct birational models of $X$, we can sensibly
define $H_2(X)$ across the whole moduli space. When we do this for a
Riemann surface $C_i$ which has positive area in the neighbourhood of the large
radius limit of one model $X_1$, the same Riemann surface $C_i$ may have
negative area near the large radius limit of some other model $X_2$.
This happens for the $\P^1$'s which are flopped when passing between
these models \cite{OP:flop}. Thus what we consider to be positive or
negative area depends on which $X_i$ we use as our starting point.

 One of the
results strongly  indicated (but not fully proven) by the
present work is that for {\it any\/} point $(t_1,\ldots,t_n)$
representing a conformal field theory,
the associated areas are non-negative when we calculate them using the \smm\
for
a suitable choice of $X_i$.
{\it In other words,
every conformal field theory in the enlarged K\"ahler moduli space
has non-negative areas
with respect to the large radius definition of size
specified by at least one of the smooth birational models of $X$
(and the method of continuation given above).}
This is the resolution of the apparent conflict
mentioned above that we put forth here (and is pictorially illustrated
later
in figure~\ref{fig:mush}).
Notice that this representation of the enlarged K\"ahler
moduli space still has all of the phases which string theory instructs
us to include
 (thereby enlarging the classical K\"ahler moduli space of
a single smooth \CY\ manifold) but that on the union of these phase
regions, the areas are constrained to be larger than certain
minimum values
(thereby reducing the classical K\"ahler moduli space).

We note that the first evidence for this conclusion in the \CY\ context
can be extracted from \cite{CDGP:}. Following \cite{AGM:II,W:phase},
the analog of figure 1 for the enlarged K\"ahler moduli space on
the quintic threefold is a $\P^1$ divided into two cells by the equator,
with north and south poles removed.
This can be thought of as arising from
$\cM = H^2(X,\C)/H^2(X,\Z)$
in the natural exponential coordinates \cite{AGM:I,W:phase}
where, as dictated by string theory, we place
no restriction on the one dimensional imaginary part of this
expression. The description of  \cite{AGM:II,W:phase} then shows that
the upper hemisphere (including arbitrary positive imaginary values in
$\cM$) corresponds to the smooth \CY\ phase while the lower hemisphere
(including arbitrary negative imaginary values in $\cM$) corresponds
to the Landau-Ginzburg phase. The analysis of \cite{CDGP:}, however,
shows that if we use the \sm\ definition of size (based on
analytically continuing the K\"ahler form from
the smooth \CY\ region as indicated above), there is a positive lower
limit on the size for all conformal theories in this enlarged
moduli space. In the present work we extend this notion to more
complicated moduli spaces which exhibit many qualitatively new
features.
Some of the regions of the moduli space we will explore can also be
described in terms of classical ideas of general relativity. We will
compare the classical version with the stringy description obtained in
this paper.

In section \ref{s:ls} we will review the local structure of the moduli
spaces of interest to this work.
 This analysis will tell us how to describe the K\"ahler moduli
space in terms of the algebraic structure of the underlying conformal
field theory --- effectively by using mirror symmetry. In section
\ref{s:gs} we will look at the global structure of the resulting
moduli space.
The discussion here will complement that of \cite{AGM:II}\ in which
toric methods were used to describe the enlarged K\"ahler moduli space
(and by mirror symmetry complex structure moduli space as well).
Here our discussion will also use toric methods, but will naturally
originate  in complex structure moduli space. In particular,
we will see that the discriminant locus (which may be
thought of as the subspace of ``bad'' conformal field theories) is closely
related to a fan structure which in turn provides data for a natural
compactification of the moduli space.

In section \ref{s:coord} we will
discuss various ways of defining the ``size'' of
a conformal field theory. Mathematically, this amounts to
 putting coordinates on the enlarged K\"ahler  moduli space
to determine a way of measuring areas at each point of the moduli space.
 It will be seen that two notions of area measurement arise.
The first notion comes from the natural coordinates that were put on
the moduli space in its algebraic  toric construction.
As we shall mention, these are also the coordinates which naturally arise
in the $N$=2 supersymmetric gauge theories employed in \cite{W:phase}.
 The other method of
area measurement comes  directly from the K\"ahler form of
the \sm\ as sketched above.

The main quantitative portion of the present work concerns presenting
methods for
the calculation of this \smm\ in section \ref{s:meas}
for various boundary or limit
 points of the enlarged K\"ahler moduli space. By studying these extreme
points in the moduli space we anticipate that our calculations will be
 sensitive to the
extreme values of volumes that can physically arise. In
section \ref{s:conc} we discuss the consequences of these
calculations and present concluding remarks.

\section{Local Structure of the Moduli Space}   \label{s:ls}

In much of what follows in both this and subsequent sections, we will
use the tool of mirror symmetry and freely interchange one perspective
with that of the mirror. To avoid confusion when we do so, let us state
our notation clearly at the outset. Let $X$ and $Y$ be a mirror pair of
\CY\ manifolds.
The mirror map takes (chiral,chiral)-fields into (antichiral,chiral)-
fields and vice versa. For both $X$ and $Y$, we will associate
deformations of the complex structure with deformations of the ring of
(chiral,chiral)-fields, and thus associate
deformations of the K\"ahler form with
deformations of the ring of (antichiral,chiral)-fields. Since we
ultimately wish to focus on deformations of the K\"ahler form of $X$
in the later
sections we use $x^i$ to denote an (antichiral,chiral)-field in the
$X$ model and $y^j$ to denote a (chiral,chiral)-field in the same
model. These are reversed by the mirror map for the $Y$ model.
This notation
is summarized in table \ref{tab:c}.

\begin{table}[h]
\begin{center}
\begin{tabular}{|c|c|c|c|}
  \hline
  Deformations&Type&$X$&$Y$\\
  \hline
  K\"ahler form&(a,c)&$x^i$&$y^j$\\
  Complex structure&(c,c)&$y^j$&$x^i$\\
  \hline
\end{tabular}
\end{center}
\caption{Notation for
the fields generating the deformations of}
\centerline{the mirror pair of
\CY\ manifolds $X$ and $Y$.}
\label{tab:c}
\end{table}

We begin with the nonlinear \sm\ given by embeddings,
$u:\Sigma\to Y$, of a
Riemann surface $\Sigma$ into a compact K\"ahler manifold $Y$ of
complex dimension 3:
\begin{equation}
  S = \frac i{4\pi\alpha^\prime}\int\left\{g_{i\bar\jmath}
	(\partial u^i\bar\partial u^{\bar\jmath}+
	\bar\partial u^i\partial u^{\bar\jmath})
	-iB_{i\bar\jmath}(\partial u^i\bar\partial u^{\bar\jmath}-
	\bar\partial u^i\partial u^{\bar\jmath})
	\right\}\,d^2z,         \label{eq:sm}
\end{equation}
where $u^i$ are holomorphic coordinates on $Y$ pulled back to $\Sigma$ and
$g_{i\bar\jmath}$ are the components of the pull-back of the K\"ahler
form.
The $B$-field is a closed real 2-form on $Y$. We will assume
that $h^{2,0}(Y)=0$ so that the cohomology class of any closed 2-form $B$
can be represented as a $(1,1)$-form
$B=\ff i2B_{i\bar\jmath}\,du^i\wedge
du^{\bar\jmath}$.
(In (\ref{eq:sm}) we also use $B_{i\bar\jmath}$ to indicate the pull-back to
$\Sigma$ of the components of $B$.) The
extra degrees of freedom introduced by the $B$-field appear to be
essential to fully understand the structure of the moduli space of
this \sm.

This \sm\ may be made into an $N$=2 field theory by introducing for each
$i$ a chiral
superfield (in both the left-moving and right-moving sense)
on $\Sigma$, $x^i$, whose lowest component is $u^i$.
Introducing superspace coordinates $\theta^\pm$ for left-movers and
$\bar\theta^\pm$ for right-movers one can show that \cite{N2sm:} the
following action
\begin{equation}
  S = \frac1{4\pi^2\alpha^\prime}\left\{\int \cK(x^i,x^{\bar\jmath})\,
	d^4\theta d^2z + 2\pi i\int_\Sigma u^*(B)\right\} \label{eq:sm2}
\end{equation}
yields (\ref{eq:sm}) as its bosonic part if
$g_{i\bar\jmath}=\frac{2i}{\pi}\frac{\partial^2\cK}{\partial
x^i\partial x^{\bar\jmath}}$.
$\cK$ is a real symmetric function of $x^i$ and
$x^{\bar\jmath}$, defined only locally on the target space.

The field theory given by (\ref{eq:sm2}) is not necessarily conformally
invariant. The conditions for conformal invariance can be studied
\cite{CHSW:,Cal:sm} by means of \sm\ perturbation theory where one
assumes that $\alpha^\prime/R^2\ll1$, where $R$ is some characteristic
radius of $Y$. This condition is called the ``large radius limit'' and
its precise meaning should become clear later in this paper. To
leading order in $\alpha^\prime/R^2$ one finds that conformal
invariance can be achieved if $B$ is harmonic and $g_{i\bar\jmath}$ is
Ricci-flat. Thus $Y$ is a \CY\ manifold.

Given any large radius \CY\ manifold we can therefore associate to it
a conformal field theory given by (\ref{eq:sm2}).
The chiral fields $x^i$ of this theory have simple multiplication properties
\cite{VW:} since one is free to make na\"\i ve definitions such as
\begin{equation}
  (x^i)^2(z) = \lim_{w\to z}x^i(w)x^i(z).  \label{eq:cr1}
\end{equation}
This simple structure for the algebra of the fields $x^i$ is the key
to being able to use the algebraic methods later in this paper.
In this paper we will often abuse notation and use $x^i$ to represent
what is really its lowest component $u^i$. This is because the
algebraic structure in the conformal field theory given by
(\ref{eq:cr1}) directly translates into a statement about complex coordinates
in algebraic geometry.

The moduli space of these
theories can now be analyzed locally as was done in
\cite{Cand:mod}. The key point is that the moduli space naturally
splits into a product of two factors. Deformations of the metric,
$g_{i\bar\jmath}$ can be divided into two types. Firstly there are
deformations of the complex structure of $Y$. These do not preserve
the (1,1)-type of $g_{i\bar\jmath}$ and introduce $g_{ij}$
and $g_{\bar\imath\bar\jmath}$ components. Such deformations form a
moduli space of complex dimension $h^{2,1}(Y)$. The deformations of
the metric that preserve its (1,1)-type can be combined with deformations
of $B$ to form the other factor in the moduli space.  This part of the
moduli space can be regarded as the set of
``complexified K\"ahler forms'' $B+iJ\in H^{1,1}(Y)$, where
$J$ is the K\"ahler form associated to $g_{i\bar\jmath}$, and it has
complex dimension
$h^{1,1}(Y)$. We will tend to drop the word ``complexified'' and refer
to the combination $B+iJ$ itself as the ``K\"ahler form'' on $Y$.
We will also fix our units of length so that $4\pi^2\alpha^\prime=1$.
Note then that
changing $B$ by adding an element of $H^2(Y,\Z)$ to it
will have no effect on the
field theory given by
(\ref{eq:sm2}) and so $B+iJ$ is best thought of as an element of
the quotient space $H^{1,1}(Y)/H^2(Y,\Z)$.

As an alternative to describing everything in terms of the intrinsic
geometry of $Y$, in some cases
one can embed $Y$ as a hypersurface in an ambient space with simpler
geometric properties. This will allow us to go some way to naturally
splitting the deformations of complex structure and K\"ahler form in
terms of the action. Consider the action (\ref{eq:sm2}) on some space
$Y_1$ with the addition of other terms:
\begin{equation}
  S = \int \cK(x^i,x^{\bar\jmath})\,
	d^4\theta d^2z + \int \lambda W(x^i)\,d^2\theta^+d^2z  +
	\int \bar\lambda\overline{W}(x^{\bar\imath})\,d^2\theta^-d^2z +
	2\pi i\int_\Sigma u^*(B),  \label{eq:smc}
\end{equation}
where $\lambda$ is a new chiral superfield and $W(x^i)$ is a
holomorphic function. This action also has $N$=2 supersymmetry. Since
no world-sheet derivatives of the field $\lambda$ appear we may
integrate it out from its equations of motion. Integrating out the
lowest component of $\lambda$ forces the condition
\begin{equation}
   W(u^i)=0. \label{eq:cond}
\end{equation}
Let us call the subspace of $Y_1$ given by (\ref{eq:cond}) the space $Y$.
Integrating out
the fermionic components of $\lambda$ forces the fermionic components of
$x^i$ to lie in the tangent bundle of the space $Y$ defined by
(\ref{eq:cond}). And integrating out
the highest component of $\lambda$ introduces an
extrinsic curvature term which along with the curvature of $Y_1$
produces the curvature of $Y$ much along the lines of \cite{Blau:}.
Thus one sees that the field theory given by (\ref{eq:smc}) on $Y_1$ is
equivalent to (\ref{eq:sm2}) on $Y\subset Y_1$. It follows that
 the condition for
conformal invariance of (\ref{eq:smc}) to leading order
is that the subspace $Y$ (rather
than $Y_1$) should be Ricci-flat. Indeed, one approach
\cite{VW:} (coming from the ideas of \cite{Friedan:})
to obtaining a
conformal field theory from this construction
is to put no condition on the metric on $Y_1$
and then consider the conformal field theory as the infra-red
renormalized limit of (\ref{eq:smc}).

In many cases all of the deformations of the complex structure of $Y$
can now be considered as deformations of the function $W(x^i)$ rather
than of the metric on the ambient space. Since this is an algebraic
question we have simplified the problem. In general one might have
some deformations of complex structure which cannot be expressed as
deformations of $W(x^i)$ \cite{GH:pdm} and we will indeed be treating
examples where this does happen. In such a case we will ignore those ``extra''
deformations and so we will only really be treating a slice of the moduli
space. There are examples known \cite{CDLS:}
where a topological class of a \CY\
manifold can be treated by more than one model of the form
(\ref{eq:smc}). It can turn out \cite{BGH:}
that in one model some deformations of complex
structure can be thought of deformations of $W(x^i)$ whereas in
another model the same deformations cannot. Because of this fact one
would expect that there is nothing special about the deformations we
are ignoring and that we should be able to see all the salient
properties of the moduli space by just looking at the slice of
$W(x^i)$-type deformations.

Consider the case where $Y_1$ is a complex projective space with, say,
the Fubini-Study metric. The infra-red limit of the
action (\ref{eq:smc}) describes a conformal \sm\ on the projective variety $Y$.
Note however that the $x^i$'s are affine rather than homogeneous
coordinates on $Y_1$. It was shown in \cite{GVW:} that a change of
variables can absorb $\lambda$ into the superpotential $W(x^i)$ and
turn the affine coordinates into homogeneous coordinates. Such a
change of variables also produced a discrete group of identifications
such that the action (\ref{eq:smc}) is an orbifold of the equivalent
action written in homogeneous coordinates.
 Similar results are also
obtained when $Y_1$ is a weighted projective space (or even a more
general toric variety) and the resulting
$x^i$ coordinates are quasi-homogeneous coordinates. From now on, to
improve notation, we will rewrite the coordinates $x^i$ as $x_i$.
Since these will always be coordinates in some flat affine space
(of which the weighted projective space or toric variety is
a quotient \cite{Cox:}), no confusion should arise.

Recently, Witten \cite{W:phase}\ has analyzed Calabi-Yau \sm s and
their relationship to Landau-Ginzburg theories. This analysis has played
a crucial role in understanding the phase structure of these theories as
discussed in our introductory remarks. It also helps to clarify why
algebraic methods suffice for understanding particular sectors of
moduli space, as we now indicate.

In Witten's approach, one begins with the action for an $N$=2 supersymmetric
two dimensional quantum field theory with a nontrivial gauge group, which
for ease of exposition we temporarily take to be $U(1)$. The action for
this theory is
\begin{equation}
  S=\int f_{\rm kin}(x_i,y_j)\,d^4\theta d^2z
	+t\int f_{\rm FI}(y_j)\,d\theta^+d\bar\theta^-d^2z
	+\int W(x_i)\,d^2\theta^+d^2z + \hbox{h.c.},
   \label{eq:LG}
\end{equation}
where $f_{\rm kin}(x_i,y_j)$ and $f_{\rm FI}(y_j)$ (the
Fayet-Illiopoulos $D$-term) are functions
which we will not concern ourselves with in this paper.

One can then study this theory for various values of the parameter $t =
r + i \theta$.
As shown in \cite{W:phase}, for $r$ large and positive, this theory
is a \sm\  on the \CY\ space given by $W = 0$ in a suitable weighted
projective space. For $r$ large (in absolute value) and negative,
the theory is interpretable as an orbifold of a Landau-Ginzburg
theory with superpotential
$W$. In the infra-red limit, these quantum field theories are expected to
become
conformal sigma models and conformal Landau-Ginzburg theories, respectively.
Mathematically, the physical construction
just reviewed corresponds to building various target spaces via
symplectic quotients  \cite{W:phase}. The parameter $r$ can
then be interpreted as setting the size, or more precisely, the K\"ahler
form on the resulting space. In more general examples \cite{W:phase},
the number of $t$ parameters equals the dimension of $H^{1,1}$ of
the associated
\CY\ space.\footnote{More precisely, the number of $t$'s equals the
dimension of that part of $H^{1,1}(Y)$ which arises from the ambient
variety $Y_1$.}
One of the results of the present study is to make geometrical sense
of such ``K\"ahler forms'' which a superficial analysis suggests will become
negative on part of the
parameter space.
We will return to a discussion of the $t$ coordinates and these issues
shortly.

As is well known \cite{Cand:coup,VW:,LVW:,W:phase}, at the conformal
limit,
some of the equations of motion of (\ref{eq:LG}) yield
\begin{equation}
  \frac{\partial W}{\partial x_i} = 0. \label{eq:eom}
\end{equation}
The important point for our purposes is that if we assume that all the
deformations of the complex structure of $Y$ are encoded in
the function $W(x_i)$, we can study the complex structure moduli space
using algebraic methods. Namely,
the fields $x_i$
obey the multiplication rules of the chiral ring \cite{VW:}
\begin{equation}
  \cR = \frac{\C[x_0,x_1,\ldots]}{\left(\frac{\partial W}{\partial
x_0},\frac{\partial W}{\partial x_1},\ldots\right)},
\end{equation}
where $(I_1,I_2,\ldots)$ represents the ideal generated by $I_1,I_2,\ldots$.
In the case that $Y$ is 3-dimensional, this ring encodes much of the
information concerning the 3-point functions in the conformal field
theory.

Because the $x_i$ are (quasi-)homogeneous coordinates, or equivalently
because they are charged under the $U(1)$ symmetries of the $N$=(2,2)
algebra, the ring $\cR$ is graded. Elements of the ring with left and
right charge (1,1) may be added to $W(x_i)$ in the action (\ref{eq:LG})
to give another valid theory. Such fields thus form truly marginal
operators.

We will now attempt to describe the deformations of K\"ahler form in
the same language. We will begin by describing the deformations of the
complex structure of a \CY\ threefold
$X$ by describing $X$ as the zero locus of a holomorphic function
$V(y_j)$ in some ambient space.
(This $X$ will eventually turn out to be the mirror partner of the $Y$
above, which is why we have switched to $y_j$ to denote the
(chiral, chiral) fields.)
To be concrete let us focus on
the example given by
\begin{equation}
  V = y_0^3 + y_1^3 + y_2^6 + y_3^9 + y_4^{18}. \label{eq:LGe}
\end{equation}
This example will be used repeatedly throughout this paper to
illustrate various points although, as will be apparent, the key
results are general.
By the arguments of \cite{VW:,GVW:,Martinec:}\footnote{This
proof of equivalence of
minimal models and Landau-Ginzburg theories is at the level of the
chiral ring which is all that we require in this paper. For issues
about whether such theories are completely equivalent see \cite{W:LG}.}
this corresponds to the Gepner
model $k=(1,1,4,7,16)$ \cite{Gep:}.
There is a 76-dimensional vector space in $\cR$ of fields
we can add to this action as marginal operators. The space is
generated by fields such as $y_2^3y_4^9$, $y_0y_3y_4^{10}$, etc.
When moving to affine coordinates the
Landau-Ginzburg theory is orbifolded by the $\Z_{18}$ action
\begin{equation}
  g:[y_0,y_1,y_2,y_3,y_4]\mapsto[\alpha^6y_0,\alpha^6y_1,
	\alpha^3y_2,\alpha^2y_3,\alpha y_4],\qquad
	\alpha=e^{2\pi i/18}.
\end{equation}
When we orbifold the conformal field theory by this action we expect
to obtain a point somewhere in the moduli space of theories of
\sm s on the hypersurface $X$ given by the zero locus of (\ref{eq:LGe})
in the weighted projective space $\P^4_{\{6,6,3,2,1\}}$. This orbifold
theory gives 3 twisted truly marginal operators in superfields of
charge (1,1) that
represent 3 deformations of complex structure of $X$ that cannot be given in
terms of $V(y_j)$.

Further analysis of the resulting orbifold also yields more truly
marginal operators, this time in superfields
with charge ($-1$,1). There are 7 of
these. Analysis of the Gepner model shows that 5
of these can be written in the following form:
\begin{equation}
  x_0x_1x_2x_3x_4,\; x_2^3x_4^9,\; x_3^3x_4^{12},\; x_3^6x_4^6,\;
	x_2^3x_3^3x_4^3, \label{eq:tmo}
\end{equation}
where $x_i$ is a superfield on $X$, this time {\it antichiral\/}
 in the left sector but
chiral in the right sector, with the same $U(1)$ charges as
$y_i$ except that the left-moving charge's sign is reversed. Thus
if we use the notation $\SLG$ for the Landau-Ginzburg action
(the action at the Gepner point)
we can represent
deformations of this action by
\begin{equation}
  S = \SLG + \int V_1(y_j)\,d^2\theta^+d^2z +
	\int W(x_i)\,d\theta^+d\bar\theta^-d^2z + \hbox{h.c.},
     \label{eq:defs}
\end{equation}
where $V_1(y_j)$ is a linear combination of the 76 marginal operators given by
monomials in $y_j$ and $W(x_i)$ is a linear combination of the fields
in (\ref{eq:tmo}). This gives a (76+5)-dimensional slice of the
(79+7)-dimensional complete moduli space.

These marginal operators written as polynomials in $x_i$
represent deformations of the K\"ahler form as was shown in
\cite{Hub:MKT}. Thus having formed an algebraic structure to describe
the moduli space of complex structures by embedding $X$ in some
ambient space, by going to the Gepner point in moduli space we see
a similar structure on the moduli space of
K\"ahler forms. This property is of course being generated by mirror
symmetry. As shown in \cite{GP:orb} one can take an orbifold of the
Gepner model to reverse the sign of right-moving $U(1)$-charge; in the
present formulation, this amounts to
exchanging the geometrical
r\^oles of $x_i$ and $y_i$ in (\ref{eq:defs}). The orbifold
required is a quotient by the group $(\Z_3)^3$ generated by
\begin{equation}
\eqalign{
 [y_0,y_1,y_2,y_3,y_4]&\to[\omega y_0,y_1,y_2,y_3,\omega^2y_4]\cr
 [y_0,y_1,y_2,y_3,y_4]&\to[y_0,\omega y_1,y_2,y_3,\omega^2y_4]\cr
 [y_0,y_1,y_2,y_3,y_4]&\to[y_0,y_1,\omega y_2,y_3,\omega^2y_4],\cr}
	\label{eq:morb}
\end{equation}
where $\omega=\exp(2\pi i/3)$. Indeed, of the 76 monomials giving
deformations of $V(y_j)$, the only ones invariant under
(\ref{eq:morb}) are obtained from
the 5 monomials in (\ref{eq:tmo}) by replacing $x$ by $y$.

Thus we arrive at the conclusion that we can study (part of)
the K\"ahler moduli
space of the \CY\ space $X$ corresponding to the hypersurface given by the
zero locus of (\ref{eq:LGe}) in $\P^4_{\{6,6,3,2,1\}}$ by considering
an orbifold of the theory given by
\begin{equation}
  S = \SLG + \left(\int W(x_i)\,d^2\theta^+d^2z +
	\hbox{h.c.}\right). \label{eq:LGp}
\end{equation}

\section{Global Structure of the Moduli Space}  \label{s:gs}

In this section we shall describe the global structure of
the enlarged moduli space of
K\"ahler forms on the \CY\ space $X$.
We did this in some detail in \cite{AGM:II} by using toric methods
and a particular construction of the so called secondary fan. In the
following we shall study this moduli space using a complimentary
approach which focuses on the complex
structure moduli space of $Y$, to which it is isomorphic by
mirror symmetry.
We will freely interchange the words ``K\"ahler moduli space of $X$''
with ``complex structure moduli space of $Y$'', via this isomorphism.

We will consider the function
\begin{equation}
\eqalign{W=a_0 x_0x_1x_2x_3x_4 +
	a_1 x_2^3x_4^9 &+ a_2 x_3^6x_4^6 + a_3 x_3^3x_4^{12} +
	a_4 x_2^3x_3^3x_4^3\cr
	&+a_5x_0^3+a_6x_1^3+a_7x_2^6+a_8x_3^9+a_9x_4^{18} = 0.\cr}
	\label{eq:gen}
\end{equation}
If we put $a_5=a_6=\ldots=a_9=1$ then we recover the superpotential of
(\ref{eq:LGp}) and we may use the 5 complex numbers $a_0,\ldots,a_4$
to parameterize the moduli space of K\"ahler forms on $X$. In this
paper however we are particularly interested in the {\em global\/}
form of the moduli space and the act of setting $a_5=a_6=\ldots=a_9=1$
would exclude certain limit points from our moduli space.

Given the fact that the scaling $x_i\to\lambda_i x_i$ is nothing more
than a reparametrization of the theory one can immediately see that we
have a $(\C^*)^5$ group of symmetries of this family of theories.
Actually in this example this
$(\C^*)^5$ is the maximum possible connected
 group of reparametrization symmetries ---
a fact which is important in this analysis. See \cite{AGM:mdmm} for
a discussion of this point.\footnote{Note that we have left open
the possibility that the full group of reparametrization symmetries
is not connected; in that case, in order to form the true moduli space
we would need to mod out by an additional finite group action, the
action of the group of connected components.  We suppress consideration
of that action in what follows.}
If we initially impose the
condition that $a_0,a_1,\ldots,a_9\neq0$ then the $a_k$ coordinates
naturally span $(\C^*)^{10}$. The $(\C^*)^5$ group of symmetries acts
without fixed points on this space and so part of our moduli space is
the space $\cMu\cong (\C^*)^5$ defined by
\begin{equation}
\cMu =  \frac{(\C^*)^{10}}{(\C^*)^5} . \label{eq:mod0}
\end{equation}
Note that $\cMu$ is constructed by modding out {\em
fully\/} by the $(\C^*)^5$-action. Setting $a_5=a_6=\ldots=1$ for
example would not be enough since it still leaves a residual
$\Z_{18}$ group of reparametrization symmetries.
This is in fact the origin of the ``extra'' discrete symmetries of moduli
spaces which have often been encountered in explicit examples
\cite{CDGP:,Mor:math,Mor:PF,AGM:I,CDFKM:I}.

We have excluded from this space $\cMu$ all points where any of the
$a_k$'s vanish. So, for example, we have omitted the Fermat point
(i.e., the form in (\ref{eq:LGe})).
On the other hand, we have implicitly included points at which the
hypersurface defined by (\ref{eq:gen}) acquires extra singularities,
and such points do not belong in the moduli space.  Our strategy now is
to enlarge $\cMu$ to a compact space $\cMc$, and then to analyze
the locus within $\cMc$ which corresponds to the set of ``bad'' conformal
field theories.  Removing that locus from $\cMc$ would then produce
the actual moduli space.

Adding in points to compactify $\cMu$ to a space $\cMc$
is far from a unique process. The study of compactifications of
$(\C^*)^n$ is known as {\em toric geometry}. One describes the data
of the compactification in terms of a fan of cones in $\R^n$
where each cone has a polyhedral base and has its apex at $O\in\R^n$.
In \cite{AGM:II} it was shown that the set of cones one naturally uses
to compactify (\ref{eq:mod0}) are given by some generalized notion of
the K\"ahler cones of $X$ and its relatives. In this section we will
motivate this collection of cones in a different manner --- namely in
terms of the natural structure of the complex structure moduli space
of $Y$.

\subsection{The Discriminant}

For fixed values of $a_0$, \dots, $a_9$,
the zero locus of (\ref{eq:gen}) defines a hypersurface $Y$ in a toric
variety.  This toric variety can be represented as an orbifold of
$\P^4_{\{6,6,3,2,1\}}$ by the group (\ref{eq:morb}), or it can be
represented more directly through toric constructions as discussed
in \cite{Batyrev1:,AGM:II}.
Consider the case
that there is a solution to the set of equations
\begin{equation}
  \frac{\partial W}{\partial x_i} = 0,\qquad\forall i.  \label{eq:d1}
\end{equation}
(This should be contrasted to (\ref{eq:eom}) which is a statement
about the {\em operators\/} $x_i$. (\ref{eq:d1}) is a statement about
the {\em complex numbers\/} $x_i$.) If this condition holds for
some point $p\in Y$ (but not for all points in $Y$)
then $Y$ will be singular at $p$. If
(\ref{eq:d1}) has no solution then $Y$ is smooth (except for quotient
singularities inherited from the ambient toric variety).

Clearly the condition that (\ref{eq:d1}) has a solution is an
algebraic problem and should be expressible in terms of a condition on
the coefficients $a_k$. The locus of points
satisfying this condition form a subspace
in $\cMc$ which is called the ``discriminant locus''.

If one tries to construct a conformal field theory corresponding to a
point in the discriminant locus one runs into difficulties. When $Y$
is smooth, the chiral ring $\cR$ is well-behaved in the sense that it
is generated as a vector space by a finite number of elements. These
elements correspond to the chiral primary fields of the conformal
field theory. When one moves onto the discriminant locus, the chiral
ring ``explodes'' in the sense that it now appears to give an infinite
number of chiral primary fields. When one tries to use the ring to
calculate 3-point functions one also runs in to trouble. Indeed if one
tries to associate a conformal field theory to such a point one
appears to demand that at least some 3-point functions are infinite.
Thus, the discriminant locus may be thought of as the subspace of
``bad'' theories. It may be that there is some way of taming such
theories, indeed many of the points we will consider which are added
to $\cMu$ to form $\cMc$ will be in the discriminant locus and we will be
able to remove the infinities. For points on the discriminant locus
within $\cMu$ however one must resolve questions such as the conformal
field theory description of the conifold transitions of \cite{CGH:con}
and such conformal field theories would appear to be necessarily badly
behaved in some sense.

For all but the simplest examples, the discriminant locus is very
complicated. In our example we will not be able to calculate the full
discriminant but we will be able to obtain much of the information we
need to study the global structure of the moduli space. The method we
will follow is that presented in \cite{GZK:d}.

First let us look at the condition that (\ref{eq:d1}) has a solution
for $x_0,x_1,\ldots,x_4\neq0$. This can be written in the form
\begin{equation}
  \Delta_0(a_k) = 0,    \label{eq:d2}
\end{equation}
where $\Delta_0$, called the {\em regular discriminant},
 is some polynomial function of the $a_k$'s. The regular discriminant
locus thus obtained is part of the discriminant locus we want within
$\cMc$. The parts we have missed are, of course, the points for which
(\ref{eq:d1}) is satisfied only when at least one of the $x_i$'s vanish.

The work of \cite{GZK:d} then proceeds as follows. First we need to
introduce the {\em Newton polytope\/} for (\ref{eq:gen}). This was
done in \cite{AGM:II} but we will repeat the main points here.
Consider representing the monomial $x_0^{n_0}x_1^{n_1}x_2^{n_2}
x_3^{n_3}x_4^{n_4}$ by the point $(n_0,n_1,n_2,n_3,n_4)$ in $\R^5$.
The equation (\ref{eq:gen}) can thus be represented by a set of 10
points in $\R^5$. Call this set of points $\cA$.
These points lie in a hyperplane in $\R^5$ and in a 4-dimensional
polytope whose corners are
defined by the monomials with coefficients $a_5,a_6,a_7,a_8,a_9$. Call
this polytope $P^\circ$. We can define a lattice $N$ within this
$\R^5$ such that $\cA = P^\circ\cap N$.

For each face, $\Gamma$, of this polytope
(of any codimension, including codimension
zero) we can define another
equation given by the points in that face. For example, one
of the codimension 1 faces corresponds to
\begin{equation}
W_\Gamma = a_2 x_3^6x_4^6 + a_3 x_3^3x_4^{12} +
	a_5x_0^3+a_6x_1^3+a_8x_3^9+a_9x_4^{18} = 0. \label{eq:egG}
\end{equation}
This defines another Newton polytope and we can define the regular
discriminant related to it.
For the face $\Gamma$ given by (\ref{eq:egG}), we would define this regular
discriminant $\Delta_0^\Gamma$ in terms of the condition that
all
$\partial W_\Gamma/\partial x_j=0$ for some $x_j$ all nonzero where the index
$j$ runs over the set $\{0,1,3,4\}$. This is similar to the part of
the discriminant we missed with the regular discriminant when $x_2=0$. We have
to be careful about the fact that the full discriminant required the
condition that $\partial W/\partial x_2=0$ whereas this was not
required for $\Delta_0^\Gamma$. Actually this doesn't matter. Setting
$x_2=0$, we have
\begin{equation}
  \frac{\partial W}{\partial x_2} = a_1x_0x_1x_3x_4\label{eq:ds1}
\end{equation}
but we also have
\begin{equation}
  \frac{\partial W_\Gamma}{\partial x_0} = 3a_5x_0^2. \label{eq:ds2}
\end{equation}
Thus, in the definition of $\Delta_0^\Gamma$, where the vanishing of
(\ref{eq:ds2}) is
imposed we obtain $x_0=0$ but this forces (\ref{eq:ds1}) to vanish.
Thus $\Delta_0^\Gamma$ does represent the discriminant of $W$ when
$x_2=0$ and $x_j\neq0$.

We can now define the {\em principal discriminant\/} as
\begin{equation}
  \Delta_p = \prod_{\Gamma\subseteq P^\circ} \Delta_0^\Gamma,
\end{equation}
where $\Gamma$ ranges over all faces of $P^\circ$ from $P^\circ$
itself to just the vertices of $P^\circ$. We wish to declare that the
condition $\Delta_p=0$ is precisely the condition that the associated
quantum field theories are bad. From the reasoning given for the
example when $\Gamma$ is given by (\ref{eq:egG}) this is true for all
points in $\cMu$. When we compactify $\cMu$ to form $\cMc$, parts of
the principal discriminant locus $\Delta_p=0$ will coincide with parts of the
divisor added to compactify $\cMu$. Whether such conformal field
theories are bad would appear to rest on precise definitions of
``badness''. We will elucidate this point by examples below.

The methods of \cite{GZK:d} can now be used to give information about
$\Delta_p$.
Actually we will not be able to construct all of $\Delta_p$ but we
will be able to calculate the key parts. For what we mean by ``key
parts'' we will now turn to a description of the asymptotic behavior
of the discriminant.

The principal discriminant $\Delta_p$ is a complicated polynomial in
the variables $a_k$. As we wonder around the compactified
moduli space $\cMc$ we
encounter regions where there is
one particular monomial $\delta_\xi$ within the
polynomial $\Delta_p$ whose modulus is much bigger than the modulus of
any other monomial. We can map out the general form of such regions as
follows. We will begin by just considering the subspace
$\cMu\subset\cMc$. Choose an explicit isomorphism
$\cMu\cong(\C^*)^5$, and let
$\tilde{a}_l\in\C^*$ be the coordinate from the $l$th factor in
$(\C^*)^5$.
(The coefficients $a_k$ in (\ref{eq:gen}) can then be expressed in terms
of the $\tilde{a}_l$, $l=1,\dots,5$.)
We make a change of moduli space parameters by
\begin{equation}
  \tilde{a_l} = e^{2\pi ib_l},\qquad l=1,\ldots,5. \label{eq:pv1}
\end{equation}
Let us also introduce a space $U\cong\R^5$ with coordinates $u_l$
given by the {\it imaginary\/} part of $b_l$, i.e., $u_l=-\frac1{2\pi}\log
|\tilde{a}_l|$. This defines a projection of the
moduli space $\pi_U:\cMu\to U$. (Later we will put $b_l=B_l+iJ_l$ in
some sense so we
expect $U$ to be the space of (real) K\"ahler forms when interpreted
in the mirror setting on $X$.)

Suppose now we consider a generic ray in $U$ that begins at the
origin, $O$, and
moves out to infinity. It is simple to see that if one is sufficiently
far out along such a ray then a single term in the discriminant polynomial
$\Delta_p$ will dominate it.
This is because the modulus of all the $a_k$ parameters will be very
large or very small, and since each monomial in $\Delta_p$ appears with
differing exponents of $a_k$'s and the ray is in a generic direction,
one monomial will contain the right exponents to win out over the
other monomials. Thus if we consider a very large $S^4$ in $U$ with
its center at $O$, then to almost every point on this sphere we can
associate a particular monomial $\delta_\xi$ in $\Delta_p$ which
will dominate.
Asymptotically as the radius of the sphere approaches infinity we can
cover $S^4$ with regions, each of which is associated to some
monomial $\delta_\xi$. Points along the boundaries of these regions,
i.e., where the regions touch will thus correspond to theories where
two or more of the dominating terms in $\Delta_p$ are (asymptotically) equal in
modulus.

The set $\{\delta_\xi\}$ of all the monomials which have some region
on the limiting $S^4$ associated to them will not, in general,
include all the terms in $\Delta_p$. There will be some terms
which never dominate $\Delta_p$ by themselves anywhere on the $S^4$.

To each element $\delta_\xi$ of our set of monomials we may take the
region in the $S^4$ at infinity described above and join all such
points to $O$ by rays. This associates a cone in $U$ to $\delta_\xi$.
The set of all such cones together with the subcones generated
by the boundaries of the regions in $S^4$ combine to form a {\em
fan\/} in $U$. This fan is the {\em secondary fan\/} that was
described in \cite{AGM:II} (although one should note that in
\cite{AGM:II} the secondary fan was described from the mirror K\"ahler
form perspective --- the equivalence of the two descriptions follows from
\cite{GZK:sp}).
The term {\em big\/} cones will be
used to denote the cones associated to the regions, as opposed to the
lower-dimensional cones arising from the boundaries between regions.
By means of the projection map $\pi_U$, this fan naturally
breaks the compactified moduli space $\cMc$ itself up into different regions.

We want to understand the transitions between regions of $\cMc$, and how
they are related to the zeros of $\Delta_p$ (i.e., to the discriminant
locus).
Let us write
\begin{equation}
  \Delta_p=\sum_\xi r_\xi\delta_\xi + \widetilde\Delta_p, \label{eq:Drd}
\end{equation}
where $\widetilde\Delta_p$ represents all the terms which do not dominate in
any big cone in $U$.
$\Delta_p$ may be normalized such that $r_\xi\in\Z$.
Although the discriminant locus has real codimension 2 in $\cMc$,
we can expect its image in $U$ to be of the same dimension as $U$ since
$U$ is half the dimension of $\cMc$.\footnote{This would fail if $U$ had
dimension $1$.}  We restrict the discriminant polynomial
to a large sphere $S^4$,
and consider the asymptotic behavior of the image of $S^4\cap(\Delta_p=0)$
under $\pi_U$ as the radius grows.
On the limiting $S^4$ ``at infinity,'' it is clear that in the interior of
each region, $\Delta_p$ cannot vanish since $\Delta_p\simeq
r_\xi\delta_\xi$. It is only when one approaches the boundary of a
region that there is a possibility of a zero in $\Delta_p$. Actually we will
argue that the image of the discriminant locus in $U$ provides codimension
one walls which asymptotically follow the walls of the big cones as
one moves out away from $O$.

Consider a point well away from $O$ in a codimension-one wall in $U$
separating
two big cones associated to $\delta_1$ and $\delta_2$. Let us assume that
this point is nowhere near any other big cones.
In this case one might at first
suspect that $\Delta_p$ will be dominated by $r_1\delta_1+
r_2\delta_2$. In most cases however some other terms from
$\widetilde\Delta_p$ will also become important. Now consider the line
in $U$ going through this point in a
direction normal to this wall.
Choose the values of the real part of $b_l$ in the directions normal to
this line.
Consider the complexification of this line to an algebraic curve in
$\cMc$ specified by these values of the real part of $b_l$.
That is, the points on this curve map to the line in $U$ and
correspond to various values of the real part of $b_l$ in the same direction.
There will
be at least one solution to $\Delta_p=0$ along this line.
As we vary the other components of the real part of $b_l$ we can move
this solution to map out a region of this line. We know however that
this image of the discriminant cannot fill up the whole of $U$ and is
actually squeezed into a real codimension one space as one approaches
the $S^4$ at infinity.
In some cases,
as we will see later, this zero in the discriminant
occurs precisely on the wall between big cones but in the
general case the image of the discriminant locus asymptotically approaches
a hyperplane parallel to the wall in question. In figure \ref{fig:as}
we show what might happen in an example where $U$ is 2-dimensional. Note the
fact that the discriminant locus carves up $U$ asymptotically into regions
given by the cones of the secondary fan except for a shifting given by
the exact form of $\Delta_p$ near this wall. Later on we will describe
these one complex dimensional subspaces of $\cMc$ for lines in $U$
infinitely far away from $O$ and we will calculate where the
discriminant locus intersects these subspaces.

\iffigs
\begin{figure}
  \centerline{\epsfxsize=11cm\epsfbox{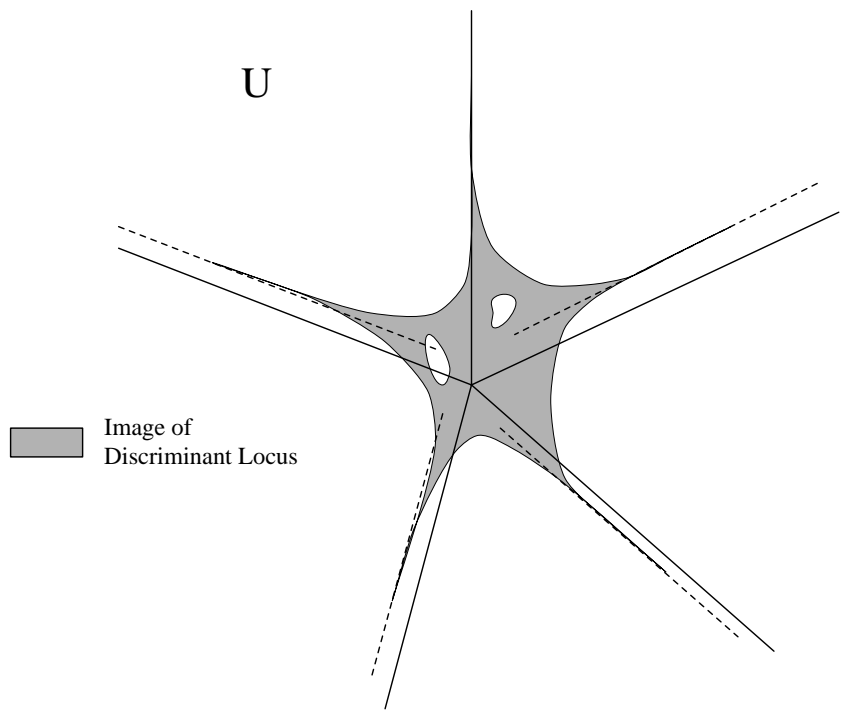}}
  \caption{The image the discriminant locus in $U$.}
  \label{fig:as}
\end{figure}
\fi

We have now arrived at the ``phase'' structure of the moduli
space described in \cite{W:phase,AGM:II}. Each big cone in the
secondary fan corresponds to a region of moduli space under the
projection $\pi_U$. As we move from one region to another there is a
singularity that one may encounter given by the discriminant locus.
Notice that in the moduli space $\cMc$, one has to aim correctly to hit
this singularity --- the discriminant locus in $\cMc$ is {\em complex\/}
codimension one and hence may be avoided.

\subsection{Compactification of the Moduli Space} \label{ss:cp}

The fan structure in $U$ may now be used to specify a
compactification $\cMc$ of $\cMu$ by the usual methods of toric
geometry. This, as we will now describe, adds in points in the moduli
space associated with points at infinity in $U$.

First we need to give a lattice structure to $U$, i.e., to specify a module
$\Z^5$ within the vector space $\R^5$.  We use our coordinates
$u_l$ specify this structure, leading to the lattice points being those with
integer coefficients
$u_l\in\Z$ for $l=1,\ldots,5$. Consider now each one-dimensional ray
$\chi$ in our
fan in $U$ and associate to it the lattice point it passes through
which is closest to $O$. Call this point
$(p_1(\chi),p_2(\chi),\ldots,p_5(\chi))$. In our example
each big cone
in the secondary fan is a so-called simplicial cone
which simply means that it is
subtended by 5 rays $\chi_1,\ldots,\chi_5$. For each big
cone let us introduce a set of coordinates $(z_1,z_2,\ldots,z_5)$
related to the coordinates $\tilde a_l$ of $(\C^*)^5$ by
\begin{equation}
\eqalign{
  z_1^{p_1(\chi_1)}z_2^{p_1(\chi_2)}\ldots z_5^{p_1(\chi_5)}
	&= \tilde a_1 \cr
  z_1^{p_2(\chi_1)}z_2^{p_2(\chi_2)}\ldots z_5^{p_2(\chi_5)}
	&= \tilde a_2 \cr
  \vdots\hskip10mm&\cr
  z_1^{p_5(\chi_1)}z_2^{p_5(\chi_2)}\ldots z_5^{p_5(\chi_5)}
	&= \tilde a_5 \cr
} \label{eq:ccs}
\end{equation}
If the $z_l$'s are all nonzero, they can be taken as coordinates in
another $(\C^*)^5$, and (\ref{eq:ccs}) defines a map $(\C^*)^5\to(\C^*)^5$
which is finite-to-one. (It will be one-to-one if the determinant of
the matrix $(\,p_i(\chi_j)\,)$ is $\pm1$.)
We add to $\cMu$ the points given by the vanishing of
any number of the $z_l$'s, i.e., we extend the $(\C^*)^5$ space given
by the $z_l$ coordinates to $\C^5$. Thus each big cone provides a
partial compactification of $\cMu$. For each big cone, the points added
locally form the structure of coordinate hyperplanes, i.e., 5
hyperplanes intersecting transversely at a point. This point of
intersection will be considered the ``point at infinity'' or
``limit point'' associated
to the big cone.

When we apply the above process for all of the cones in our fan we
form a complete compactification of $\cMu$ and this completely specifies
our compactified moduli space $\cMc$.
The points added form a divisor with normal
crossings, i.e., a codimension-one subspace in $\cMc$ whose
irreducible components intersect transversely.
It should be noted that the compactified moduli space $\cMc$ formed
this way will not, in general, be smooth and is not smooth in our
example. While one might wish to resolve the singularities in this
space to address questions about monodromy of periods \cite{Mor:cp},
in this paper it will be
important to retain the singularities. It would appear therefore, that
at least in some ways, $\cMc$, and not its resolution, forms the
most natural compactification of the
moduli space of K\"ahler forms on $X$.

There is a relationship
between the codimension
of parts of the compactification set and
the dimension of the cones in our fan. To each cone of real dimension $n$
in our fan, we can associate an irreducible (sub)space of the
compactification divisor of $\cMc$ of complex
codimension $n$. For
example, each big cone describes a point --- the ``point at infinity'' above
which is the point $(0,0,\ldots,0)$ in the
coordinates $z_l$. Each one-dimensional ray in the fan corresponds to
an irreducible component of the compactification
divisor. Of particular interest in
this paper will be the codimension one walls in the fan which give
one complex dimensional subspaces within the compactification divisor.
The fact that these one dimensional subspaces are compact and toric
means that they are {\em rational curves\/} (isomorphic to $\P^1$) in $\cMc$.

Now we will describe how to calculate $r_\xi\delta_\xi$ for each big
cone in $U$ following \cite{GZK:d}. Firstly we need to associate a
triangulation of the set of points $\cA$ to each big cone. This was
described in detail in \cite{AGM:II} and we will again review it
briefly here. The triangulation will be determined by the choice of
a ``height function'' $\psi$, which to
each point $\alpha_k\in\cA$, associates a
``height'' $\psi(\alpha_k)\in\R$. As the name suggests one should
think of this as providing the extra coordinate for an embedding
$\cA\in\R^6$. The space $U$ can then be considered to be the space of
``relative heights''. That is, fix the position of $P^\circ$ within
this $\R^6$ space by, say, fixing the heights of the vertices of
$P^\circ$ be have height zero\footnote{This is possible in our example
since the Newton polygon is simplicial; in general, normalizing the
heights is more complicated.} and then let the other points vary to
fill out a space of relative heights $\cong U$.

Now consider stretching a piece of rubber over these points which are
at various
heights. If the heights are generic,
the shape thus formed specifies a triangulation of $\cA$. The flat
faces of the shape will be simplicial specifying the simplices in the
triangulation. Points not touching the shape, i.e., below the
stretched film are not considered in the triangulation. Thus, for
example if $\psi(\alpha_k)$ is negative for all points not vertices of
$P^\circ$ and zero for the vertices of $P^\circ$, then the
triangulation consists of just the simplex $P^\circ$ itself. By
labeling the points in $U$ according to which triangulation they
give, one obtains a fan structure with each big cone specifying a
triangulation. Cones of lower dimension specify non-generic heights
where one is on the borderline between two or more triangulations.

Let us recall that we began by analyzing the principal discriminant
and finding that some of the monomials contained in this naturally
dominated in some region of moduli space. To each such monomial we
associated a big cone in a fan in the space $U$. Now we have
associated a triangulation of the point set $\cA$ to each such cone.
Now we can state the algorithm from \cite{GZK:d} which directly specifies
the monomial from the triangulation:

Each big simplex (i.e., simplex of maximal dimension), $\sigma$,
in the triangulation of $\cA$ can be given a normalized volume, $\Vol(\sigma)$,
proportional to its actual volume. (The constant of proportionality is
fixed so that in a maximal, or ``complete,''
 triangulation, all simplices are have
$\Vol(\sigma)=1$.) Using the notation of (\ref{eq:Drd})
\begin{equation}
  \eqalign{r_\xi &= \pm\prod_{\sigma\in T_\xi}
	\Vol(\sigma)^{\Vol(\sigma)}\cr
  \delta_\xi &=
  \prod_k a_k^{\left(\sum_{\sigma\ni\alpha_k}\Vol(\sigma)\right)},\cr}
		\label{eq:rr}
\end{equation}
where $T_\xi$ is the triangulation associated the this monomial. The
summation in the equation for $\delta_\xi$ is taken over the set of
$\sigma\in T_\xi$ such that $\alpha_k$ is a vertex of $\sigma$. The
relative signs of $r_\xi$ can also be determined by a process given in
\cite{GZK:d}. These signs will play an important
r\^ole in our analysis and we will
give an equivalent method of their calculation later in this paper.

In our example, the polytope $P^\circ$ has normalized volume 18 and
there are 100 triangulations leading to convex height functions
$\psi(\alpha_k)$. (Actually these give all possible triangulations of
the point set $\cA$ in this case.) Thus, there are 100 monomials in
the sum in (\ref{eq:Drd}) and 100 big cones in our fan in $U$. As a
simple example, the monomial given by the triangulation consisting of
one simplex ($P^\circ$ itself) we have
\begin{equation}
  r_\xi\delta_\xi = -18^{18}\,a_5^{18}a_6^{18}a_7^{18}a_8^{18}a_9^{18}.
\end{equation}
At the other extreme, there
 are 5 possible complete triangulations of the set $\cA$. One of
them has
\begin{equation}
  r_\xi\delta_\xi = a_0^{18}a_1^{8}a_2^{6}a_3^{8}a_4^{10}
	a_5^{12}a_6^{12}a_7^{6}a_8^{6}a_9^{4}. \label{eq:del1}
\end{equation}
Intermediate triangulations give terms such as
\begin{equation}
  r_\xi\delta_\xi = 1259712\,a_0^{15}a_1^{15}a_3^{10}
	a_5^{13}a_6^{13}a_7^{8}a_8^{13}a_9^{3}.
\end{equation}


\section{Putting coordinates on the moduli space and
the definition of size}  \label{s:coord}

So far we have built the complete compact space $\cMc$ which compactifies
the space of complexified
K\"ahler forms on $X$ (or equivalently, complex structure moduli of $Y$)
in the context of conformal field theory. What
we have does not, at first sight,
resemble a space of classical K\"ahler forms however. In this
section we will review how the structure of $\cMc$ is linked to the
classical notion of K\"ahler forms in some limiting sense and then
show how this linkage
may be continued to all points in the moduli space. In this way
we will extend the usual mathematical notion of volume or size from classical
to quantum geometry, as discussed in the introduction. We will explicitly
do this by defining coordinates on the K\"ahler moduli space. In essence,
the particular continuation of size from geometry to conformal field
theory depends upon how we coordinatize the K\"ahler moduli space.
In most contexts we do not place much importance on   coordinate
choices as we expect all physical conclusions to be independent of the
possible choices. This reasoning is, of course, true here, but there
is an important distinction. The space upon which we are putting
coordinates is a moduli space, i.e. a space of coupling constants for a class
of conformal field theory actions. Different choices of coordinates
correspond to different ways of representing and parametrizing these
quantum systems.  Our goal in this paper is to {\it interpret\/}
the geometrical content of all of the conformal theories in the enlarged
K\"ahler moduli space. This goal, in turn, will dictate particular
ways of representing these theories (via nonlinear sigma models) and
particular parametrizations (directly in terms of their complexified
K\"ahler forms and their analytic continuations) --- i.e. a
particular choice of coordinates on the moduli space. This, we shall
argue, is the choice which makes the geometrical interpretation most clear,
but it is certainly not unique.

In fact, we will find it useful to introduce two particular coordinate
systems on the enlarged K\"ahler moduli space of $X$ --- each of which
will give rise to a definition of ``size'' at every point of the moduli space.
For both of these, we will give an implicit
 definition of length (inferred from an explicit definition of
area) such that both of the following hold:
\begin{enumerate}
  \item The definition of length in each conformal field theory is
given in terms of the fundamental data determining the latter, i.e.
its two and three point functions.
  \item If the underlying conformal
theory is smoothly deformed to a large radius \CY\ sigma model,
then the conformal field theory definition of length asymptotically
approaches the standard geometrical definition of length on the
\CY\ space.
\end{enumerate}
It might be worth pointing out here that such definitions of length
should not be expected to be modular invariant.  For instance,
specifying that
a circle has radius $R$ in string theory is not
a modular invariant notion because the specified radius
 obviously differs (almost
everywhere) from the equivalent radius $\alpha^\prime/R$.
Even so, we are certainly justified in saying that string theory on a circle
imposes a lower bound of $(\alpha^\prime)^{1/2}$ on the
radius --- the point being
that this is true in a fundamental domain (in the Teichm\"uller space)
for the modular group.
Thus, in this case,
in order to associate a notion of size to conformal field theories,
we are obligated to make a choice of fundamental domain and
work within it. As we shall see, toric geometry provides us directly
with the moduli space itself rather than the Teichm\"uller space. We
will construct a fundamental domain
 so that the large radius limit will be an element of it\footnote{More
precisely, the large radius limit is an interior point in a partial
compactification $(\cD/\Gamma)^-$ of $\cD/\Gamma$, where $\cD$ is the
fundamental domain and $\Gamma$ is the ``\sm\ part'' of the modular
group, obtained from integral shifts of the $B$ field and the
holomorphic automorphisms of $X$ \cite{Mor:cp}.}
and so we are forming the analog of the $R>(\alpha^\prime)^{1/2}$
region in the above context.

In practice, each of the definitions of length we introduce will
rely on mirror symmetry. Namely, we have complete analytic understanding
of the complex structure moduli space of, say, $Y$. Mirror symmetry
provides us with an abstract map from this moduli space to that of
the enlarged K\"ahler moduli space of $X$. Different explicit realizations
of this map will associate different coordinates and hence definitions
of length to the underlying conformal theories in the K\"ahler moduli space
of $X$.
The first explicit realization is mathematically the simplest and amounts
to extending the ``monomial-divisor mirror map'' of \cite{AGM:mdmm}\
throughout the moduli space. The same coordinates also naturally arise
from the physical approach of \cite{W:phase} from somewhat the
opposite point of view. The second explicit realization makes  use of
the results of \cite{CDGP:} which, via the sigma model,
provides a direct
link between physical observables and classical geometry.

\subsection{The monomial divisor mirror map and the \mKf}

In section \ref{s:gs} we obtained the result that the space $\cMc$
contained 100 special points which were obtained from the 100 big
cones in our fan in $U$. It was shown in \cite{W:phase,AGM:II} that
each of these points could be related to some space that modeled $X$
in the following way. First one takes the cone in $U$ associated to
the point in $\cMc$. Then one takes the triangulation $T_\xi$ of $\cA$
associated to this cone. This triangulation can then be taken as the
base of a set of cones forming a fan $\Delta^+$ (not to be confused with
the original fan in $U$). From $\Delta^+$ one builds a toric variety
$V_{\Delta^+}$ in the same way as we constructed $\cMc$ from a fan.
The target space $X$ can then be interpreted as the critical locus of
some function $W$ within this toric variety.
When identifying these models the point $\alpha_0$ associated to the
monomial with coefficient $a_0$ in (\ref{eq:gen}) plays a special r\^ole.
Key examples are as
follows:
\begin{enumerate}
  \item When $T_\xi$ is a complete triangulation of $\cA$ one has that
$V_{\Delta^+}$ is a line bundle of some smooth toric 4-fold $V$. $X$ is then
the \CY\ manifold defined by $W=0$ at infinite radius limit. \label{i:1}
  \item When $T_\xi$ is comprised of only the simplex $P^\circ$ then
$V_{\Delta^+}$ is a point. The target space is this point but the
quantum field theory includes
some massless modes around this point. This is a Landau-Ginzburg
orbifold theory.
  \item When $T_\xi$ omits some points of $\cA$ but $\alpha_0$ is a
vertex of every $\sigma\in T_\xi$ then $V_{\Delta^+}$ has the
structure of a line bundle (in a suitable sense) over some singular
space $V$. $X$ is again defined by $W=0$ and is still at some infinite
radius limit but has quotient singularities.
  \item When $T_\xi$ has more than one simplex, $\sigma$,
 but $\alpha_0$ is not a vertex of each $\sigma$ then one has some
kind of hybrid model where at least part of $X$ is given by a
Landau-Ginzburg orbifold theory ``fibered'' over a manifold of complex
dimension one or two.
\end{enumerate}

The reader may be somewhat surprised that we began with a specific
manifold $X$ but now that we have analyzed the global structure of the
moduli space of K\"ahler forms on $X$ we have 100 geometric models all
equally as valid as $X$. This is because, as emphasized in
\cite{AGM:I,W:phase},
conformal field theory happily smooths out topological transformations
of $X$ so that our moduli space will, if complete, necessarily contain
the other types of $X$ that can be reached from the original $X$.

The key example in this case is case \ref{i:1}. This allows us to identify
which cone in $U$ corresponds to the \sm\ we began with. In our
example there are 5 complete triangulations of $\cA$ and hence 5
smooth \CY\ manifolds equally valid as starting points for this
analysis.

Picking one of these models, we take the coordinates $z_l$ in $\cMc$
from (\ref{eq:ccs})
related to the corresponding big cone.
In more physical language,
a given point in the moduli space corresponds to some abstract conformal
field theory. The coordinates $z_l$
are chosen so that the complex
structure on $Y$ is such that the resulting correlation  functions
agree with those of the associated conformal field theory.
In other words, we deduce the coupling constants for the sigma model
action on $Y$ (the coefficients in \ref{eq:gen}) by ``measuring'' scattering
amplitudes (calculating correlation functions) in the chosen conformal
theory. There is no problem in carrying out this procedure since we
can calculate three point functions associated with complex structure
moduli exactly by using the results of \cite{Cand:coup,DG:exact}.
So much for the complex structure sector of $Y$.

We now state that the
complexified K\"ahler form on $X$ is given asymptotically by the {\em
monomial divisor mirror map\/} \cite{AGM:mdmm}:
\begin{equation}
  B_l + iJ_l = \frac1{2\pi i}\log(\pm z_l),  \label{eq:mdmm}
\end{equation}
where we have defined some basis, $e_l$, of $H^2(X,\Z)$, such that
\begin{equation}
  B+iJ = \sum_l (B_l+iJ_l)e_l.
\end{equation}
We then define the cycles $C_l\in H_2(X)$ by
\begin{equation}
  \int_{C_k} e_l = \delta_{kl}.
\end{equation}
and regard a choice of $B+iJ$ as a way of specifying areas:
\begin{equation}
  \Area(C_l) = \Img\int_{C_l} (B+iJ) = \frac1{2\pi}\log|z_l|.
\label{eq:Bidef}
\end{equation}
(In fact, the ``complexified areas'' $\int_{C_l}(B+iJ)$ are also determined
by this choice.)

The sign ambiguity of $z_l$ in (\ref{eq:mdmm}) is referred to in
\cite{AGM:mdmm}\footnote{This sign problem appears to have been
ignored in \cite{BvS:}.}
and
we will fix it later in this paper.
The divisors representing $e_l$ may also be determined by the
monomial-divisor map \cite{AGM:mdmm}.
Note that the monomial-divisor mirror map is consistent with the
invariance of the theory under the transformation
$B\to B+x,\; x\in H^2(X,\Z)$, and that the origin of our
coordinate patch $z_l=0$ corresponds to $J_l\to\infty$ consistent with
this point being the large radius limit of the \CY\ manifold.

This is our first definition of coordinates. We
have constructed the complete moduli space $\cMc$ of K\"ahler forms on
$X$ and put coordinates on this space that allows us to explicitly
assign an area to $2$-cycles
at every point in
$\cMc$. We may consider that the measurement of areas
on $X$ is {\em defined\/} by the choice of cohomology class
(\ref{eq:mdmm}), and that this definition
agrees with classical geometry at large radii.
This definition, as discussed above, can be phrased in terms of
the correlation function data of the underlying conformal theory.
Therefore, this definition satisfies the two properties emphasized
in the beginning of section 4.
 {\it The measurement of areas
defined in this way will be called ``the \mKf''.} This object (or rather,
it's imaginary part)
may be used in the same way as the classical K\"ahler class $J$
to measure the areas on Riemann surfaces in $X$.  (In this case, one can also
measure the volume of $X$ itself using $J\wedge J\wedge J$, and
the volumes of divisors on $X$ using $J\wedge J$, but we will concentrate
on the area measurements, for reasons we will see shortly.)

The classical geometric significance of these coordinates is most directly
gleaned from the work of \cite{W:phase}. As we have discussed earlier
and will explain more fully in
\cite{AGM:IV},
Witten's approach is the physical manifestation of the toric methods
under discussion via the relationship between holomorphic and symplectic
quotients. The real part of the coordinates $t$ (more
generally, $t_i$) which appear in the action
(\ref{eq:LG}) are, in fact, precisely the algebraic coordinates just
defined.
That is, if one wants to connect the \mKf\ to some classical notion of
distance then the \mKf\ may be thought of as arising from
the classical K\"ahler form on the target space
of the non-conformal field theory given by (\ref{eq:LG}).

With this definition of the complexified K\"ahler class $B+iJ$,
the image of the \sm\ phase under the projection $\pi_U$ is precisely the
K\"ahler cone of $X$. If one follows a path in $\cMc$ which
moves from the large radius \sm\ on $X$ to a point $m\in\cMc$ where
$\pi_U(m)$ lies outside that cone, then $J_l$ becomes negative for some
$l$ just as one passes through the wall of the cone. That is, the
area of some Riemann surface on $X$ becomes negative at this point. Thus,
the 99 other cones in $U$ can be interpreted as a \sm\ on $X$ where
some area is negative. As mentioned
in the introduction, in the first case studied in \cite{W:phase}
of the mirror of the quintic threefold, $U$ was a line and consisted
of just two cones, i.e., two rays in either direction from $O$. When
one ray is interpreted as the K\"ahler cone of the \CY\ manifold one
sees that the other region must be interpreted as a manifold whose overall
volume is negative
and that the Landau-Ginzburg orbifold theory can be thought of as
 a \CY\ manifold with overall
volume equal to $-\infty$. Our situation is similar but now we have 99
limit points where the area of some subspace of $X$ (and perhaps the
volume of $X$ itself) is
$-\infty$.

Four of these other regions actually have  all of the associated
areas being positive
if we  interpret the situation not
in terms of  $X$ but rather in terms of a topologically
different manifold --- a
{\em flop\/} of $X$ \cite{AGM:II}.
We emphasize that we have not modified the physics in any way; we have
only reinterpreted the conformal field theory in terms of its most
natural geometrical model.
 Some of the other 95 limit points
correspond to orbifolds. In this context, the orbifold points in the
moduli space of \CY\ manifolds would normally be thought of as limit points
where some divisor, the {\em exceptional divisor}, has shrunk down to
zero volume (the reverse of blowing up). When we use the
\mKf\ however we arrive at the different conclusion that the volume of
the exceptional divisor is $-\infty$ at the orbifold point.
(In terms of areas:  every Riemann surface within that exceptional
divisor has area $-\infty$.) Actually
this shift from 0 to $-\infty$ is a recurring feature of many of the other
regions. In each case one would naturally have wanted to
interpret the conformal field theory as
having some target space in which some part of $X$ has shrunk to zero
area, but in each case the area defined by the \mKf\ is $-\infty$.
The Landau-Ginzburg orbifold model is the extreme case --- here the target
space is a point, i.e., the whole of $X$ has shrunk to zero, whereas
its algebraic areas are all $-\infty$.

Thus we have seen that the \mKf\ has its advantages and disadvantages.
It is easily defined in terms of the natural coordinate charts on
$\cMc$ and it reproduces the K\"ahler cone of $X$. What one might be
uncomfortable with however is the fact that most of the moduli space
$\cMc$ is comprised of $X$'s with negative area subspaces
 and that this definition
has a complicated (and largely only implicit) conformal field theory
representation.

\subsection{The \smm}

We will now make another attempt at defining ``size,'' this
time trying to model more closely the properties of the classical
K\"ahler form. This is done at the expense of the simplicity of the
definition in terms of the natural coordinates on $\cMc$.

One can use the action (\ref{eq:sm2}) to calculate the 3-point
function between (chiral,antichiral)-fields in the resulting quantum
field theory. This is best achieved by twisting this $N$=(2,2)
superconformal \sm\ into the so-called {\em A-model\/} topological
field theory \cite{W:tsm,W:AB}. Each field can be associated to an element
of $H_4(X)$ and the 3-point functions can in principle be calculated from
intersection theory. If to each field $\phi_l$ we associate a
divisor $D_l$, then to leading order in the large radius limit we have
\begin{equation}
  \langle\phi_l\phi_m\phi_n\rangle\sim \#(D_l\cap D_m\cap D_n).
\end{equation}
(We omit the ``\#'' symbol denoting ``degree of intersection'' from now on.)
These intersection numbers agree with those predicted by
the monomial-divisor mirror map
\cite{AGM:I,Bat:q} as explained above.
Beyond this asymptotic form of the 3-point functions at large radius
limit we may ask what happens if $X$ is near, rather than at, the
large radius limit. In this case one may expand the 3-point function
out in terms of an instanton series \cite{DSWW:}.
The instantons in
question are given by holomorphically embedded $\P^1$'s
 in $X$ and for the exact form of
this instanton series one should consult \cite{AM:} (and the
references therein) but it can be stated
roughly as
\begin{equation}
  \langle\phi_l\phi_m\phi_n\rangle= (D_l\cap D_m\cap D_n)
    +\sum_\Gamma\frac{{\bf q}^\Gamma}{1-{\bf q}^\Gamma}(D_l\cap\Gamma)
    (D_m\cap\Gamma)(D_n\cap\Gamma),
\end{equation}
where $\Gamma$ is a holomorphically embedded $\P^1$
 in $X$ and ${\bf q}^\Gamma$ is a monomial
in the $q_l$'s.
We define the
parameters $q_l$ by
\begin{equation}
  q_l = \exp\{2\pi i(B_l+iJ_l)\}, \label{eq:smm}
\end{equation}
with $B$ and $J$ coming from (\ref{eq:sm2})
so that the resulting 3-point functions appear as power series in the
$q_l$'s. This leads us to another way of defining areas
for a point in $\cMc$. That is, we determine the values of
$B_l+iJ_l=\int_{C_l}B+iJ$
required to give the correct 3-point functions when these 3-point
functions are expressed as an instanton sum, i.e., as a power series
in $q_l$. We then analytically continue this object over the whole
moduli space.
{\it We will refer to this definition of area-measurement
as ``the \smm''.}
Note that to perform the analytical continuation of the \smm\
over $\cMc$ we need to make some branch cuts in $\cMc$. We claim
that there is a natural choice and we specify this choice later.

The reason that we distinguish these two definitions of area-measurement in
this paper is because they are, in fact, different. That is, in
general, $q_l\neq z_l$ so we will need to specify which coordinates
we are using, in order to specify the measures.
{}From now on we will use the symbol $B+iJ$ to
refer to the {\em\smm\/} only.

Returning to Witten's approach to the \mKf\ outlined in the previous
section we see that when one takes the renormalization group flow
limit of the field theory given by (\ref{eq:LG}) to the conformal
field theory the \mKf\ must ``flow'' to the \smm.
After all, (\ref{eq:LG}) is describing a sigma-model.

These definitions of the \mKf\ and the \smm\ are,
as constructed, in complete agreement at the
large radius limit. Thus, with our conventions about $B+iJ$
being the \smm, we can modify (\ref{eq:mdmm}) to read
\begin{equation}
  B_l + iJ_l = \frac1{2\pi i}\left\{\log(\pm
	z_l) + O(z_1,\dots,z_5)\right\}, \label{eq:mm}
\end{equation}
i.e., we expand $\log q_l$ as a power series for small $z_l$. Actually we have
not justified the omission of a constant term in the right-hand-side
of (\ref{eq:mm}) and we will return to this point briefly later.

In \cite{CDGP:} a good geometrical way of picturing these natural
\sm\ coordinates\footnote{These are sometimes called the {\it flat
coordinates\/} in the literature.}
 in terms of the mirror theory $Y$ was introduced. If we
view $\cMc$ as the moduli space of complex structures of $Y$ then a
natural set of coordinates can be introduced via the {\em
Gau\ss-Manin connection}. That is, in the case of 3-folds we introduce
the holomorphic 3-form $\Omega$ and a set of 3-cycles $\gamma_n$. One
can then define
\begin{equation}
  B_l + iJ_l = \frac{\displaystyle\int_{\gamma_l}\Omega}
	{\displaystyle\int_{\gamma_0}\Omega}, \label{eq:period}
\end{equation}
These coordinates are independent of the
normalization of $\Omega$ and will satisfy (\ref{eq:mm}) if $\gamma_0,\gamma_l$
are suitably chosen. See \cite{Mor:math} for more information.
In \cite{CDGP:} the definition of the \smm\ via
(\ref{eq:period}) was used directly to obtain the correction
terms in (\ref{eq:mm}). That is, certain 3-cycles were found and
$\Omega$ was integrated over them. In practice this method will be
unsuited to approach the problems addressed in this paper. Instead
it was noticed in \cite{CDGP:} that these periods satisfied a
differential equation and in \cite{Mor:PF} that one could use these
differential equations to find the form of (\ref{eq:mm}) without
explicitly constructing the 3-cycles $\gamma_0,\gamma_l$.

There is an important qualitative feature of the local solutions to
these differential equations.  The cycle $\gamma_0$ has the property
that $\int_{\gamma_0}\Omega$ is regular as a function of the $z_l$.
Thus, comparing (\ref{eq:mm}) with (\ref{eq:period}),
we find
\begin{equation}
2\pi i\,\int_{\gamma_l}\Omega
= \log(\pm
        z_l)\left(\int_{\gamma_0}\Omega\right) + O(z_1,\dots,z_5)
\end{equation}
which tells us that in addition to the regular solution, there is
a solution with a $\log(\pm z_l)$ type growth for each $l=1,\dots,5$.
Moreover, all the other solutions will involve products or powers of
these log terms.  All of this is discussed in more detail in \cite{Mor:cp}.

It is worthwhile noting that whereas we had no problem in in using the
\mKf\ to measure areas and volumes of $X$ and its subspaces in the
classical way, the same is not true for the \smm. For
example defining
\begin{equation}
  \Vol(X) = \int_X J\wedge J\wedge J,
\end{equation}
one would find the value of the volume behaved in an unsatisfactory way
as one moved around the moduli space. A better definition would be
some object of the form of a correlation function $\langle
JJJ\rangle$.
Some of the properties
of the \smm\ are actually quite insidious.
In the classical
picture the K\"ahler form lives in the
linear vector space $H^2(X)$. Although we tried to mimic this in the
quantum picture by exercising great care in choosing the \smm\
coordinates, the quantum corrected moduli space is not flat and
this is reflected in some non-linearity in the structure of the \smm.
This underlies the reason why the ring structure
given by the wedge product in $H^*(X)$ is not a natural object in $\cMc$.
We will refer to this issue of non-linearity
briefly again later in the paper and we
hope to address further questions about this structure in future work.
In this paper we will only attempt to use the \smm\ to
directly measure the area of Riemann surfaces in $X$. We will also use
the classical notion that if a manifold has zero volume then any
subspace within it is also of zero volume. This is the only sense in
which we will measure volumes, as opposed to areas of Riemann surfaces.

\section{Evaluating the \smm}  \label{s:meas}

\subsection{The Hypergeometric System}

In this section we will discuss the system of partial differential
equations which allow one to find the natural \sm\ coordinates
(\ref{eq:period})
required for
the \smm. With the notation of section \ref{s:gs},
let $\alpha_k\in\cA$ have coordinates $(\alpha_{k,1},\alpha_{k,2},
\ldots,\alpha_{k,5})$ in $\R^5$.
For given values of $\beta_n$
consider the following differential operators introduced in \cite{GZK:h}:
\begin{equation}
\eqalign{
  Z_n &= \left(\sum_{k}\alpha_{k,n} a_k\frac\partial{\partial a_k}\right)
	-\beta_n\cr
  \Box_l &= \prod_{m_{l,k}>0}\left(\frac\partial{\partial
	a_k}\right)^{m_{l,k}} - \prod_{m_{l,k}<0}\left(\frac\partial{\partial
	a_k}\right)^{-m_{l,k}},\cr}
\end{equation}
where $n=1,\ldots,5$ and $l$ labels a relationship
\begin{equation}
  \sum_{k}m_{l,k}\alpha_{k,n}=0, \qquad n=1,\ldots,5. \label{eq:cond1}
\end{equation}
Now we look for a function $\Phi(a_0,a_1,\ldots,a_9)$ such that
\begin{equation}
  Z_n \Phi = \Box_l\Phi = 0, \qquad\forall n,l.  \label{eq:PF}
\end{equation}
The numbers $\beta_n$ specify how $\Phi$ transforms under the
$(\C^*)^5$ action $x_i\to\lambda_i x_i$. In \cite{Bat:var}, it was
shown that the periods in (\ref{eq:period}) satisfy (\ref{eq:PF})
for a certain choice of $\beta_n$ which we will now give.

We first need to make a special choice of coordinates on the $\R^5$
space in which the points $\cA$ live. Remember that the
(quasi-)homogeneity of (\ref{eq:gen}) means that these points lie in a
hyperplane. Let the coordinates be chosen such that $\alpha_{k,5}=1$
for $k=0,\ldots,9$ and let the coordinates of $\alpha_0$ be
$(0,0,0,0,1)$. In this basis the values of $\beta_n$ required are
$\beta_n=0$ for $n=1,\ldots,4$ and $\beta_5=-1$.

We can now give a general solution to the partial differential
equations $Z_n\Phi=0$ but first we need to say more about
$(\C^*)^5$-invariant coordinates. The $a_k$ parameters transform under
the $(\C^*)^5$-action by the condition that (\ref{eq:gen}) is
invariant. This means that for each condition of the form
(\ref{eq:cond1}) we may introduce
\begin{equation}
  z_l = \prod_k a_k^{m_{l,k}} \label{eq:z2a}
\end{equation}
which are invariant under the $(\C^*)^5$-action. The fact that we are
using the same notation $z_l$ for such invariants and the coordinate
patches on $\cMc$ in (\ref{eq:ccs}) is not an oversight --- they can
be considered the same thing as we now discuss.

One of the big cones in our fan in $U$ corresponds to the
Landau-Ginzburg orbifold model. We know that the local space around the
Landau-Ginzburg orbifold point can be parametrized by $a_0,\ldots,a_4$ and
setting $a_5,\ldots,a_9=1$. Calling the coordinates for the
Landau-Ginzburg orbifold model $z_l^{\rm (LG)}$ we thus see that for
$a_5,\ldots,a_9=1$ we can define $z_l^{\rm (LG)} = a_{l-1}$. We may
remove the $a_5,\ldots,a_9=1$ condition by multiplying $a_{l-1}$ by
the necessary powers of $a_5,\ldots,a_9$ required to achieve
$(\C^*)^5$ invariance, thus we have
\begin{equation}
  z_1^{\rm (LG)}=a_0a_5^{-\frac13}a_6^{-\frac13}a_7^{-\frac16}
   a_8^{-\frac19}a_9^{-\frac1{18}},
  z_2^{\rm (LG)}=a_1a_7^{-\frac12}a_9^{-\frac12},\ldots
\end{equation}
(Fractional powers appear here because the Landau-Ginzburg orbifold point is at
a quotient singularity in $\cMc$.)

Actually there is a technical point that should be addressed here.
The above form for the Landau-Ginzburg orbifold coordinates
reflects the fact that there is a $\Z_{18}$-quotient singularity at
this point in the moduli space. Quotient singularities in one complex
dimension are trivial is the sense that they can be removed by a
change of coordinates. Our description of $\cMc$ in terms of toric geometry
automatically removes such singularities. The $\Z_{18}$-quotient
singularity in the moduli space is actually only a $\Z_6$-quotient
singularity once this process is performed. Thus in the toric
description given one should actually use a 3-fold cover of the above
coordinates. An alternative is to modify the definition of the
coordinates $(p_1(\chi),p_2(\chi),\ldots,p_5(\chi))$ in terms of the rays
of the
secondary fan. Our original definition was in terms of the {\em
first\/} point from $O$ encountered by the ray. It turns out that by
taking the third point rather than the first for one of the rays
restores the $\Z_{18}$ singularity. (Actually this also occurs when we
construct these rays as vectors from the
{\em Gale transform\/} of $\cA$\, \cite{BFS:}.)
In what follows we assume that $(p_1(\chi),p_2(\chi),\ldots,p_5(\chi))$
has been
rescaled for one of the rays in this way.

We may now use (\ref{eq:ccs}) to
give $(\C^*)^5$-invariant $z_l$ coordinates for each big cone in our
fan. Thus for each big cone in the fan we have a set of $z_l$ coordinates
and a set of conditions (\ref{eq:cond1}) given by (\ref{eq:z2a}).

It is not difficult to show that the equations $Z_n\Phi=0$ have as a
general solution
\begin{equation}
  \Phi(a_0,a_1,\ldots,a_9) = a_0^{-1}f(z_1,z_2,\ldots,z_5),
		\label{eq:genZ}
\end{equation}
where $f$ is an arbitrary function and the $z_l$'s are any set of
$(\C^*)^5$-invariant coordinates. For each big cone we can now write
$\Phi$ in the form (\ref{eq:genZ}) and write down the $\Box_l\Phi=0$
equations.

Let us choose one of the cones corresponding to the large radius limit
of a smooth \CY\ manifold and write these differential equations down.
We can specify such a cone by specifying a complete triangulation of
$\cA$. The complete triangulations of $\cA$ are unique except for the
triangle with vertices $\alpha_7,\alpha_8,\alpha_9$. We will first
concentrate on ``resolution 1'' from \cite{AGM:I} given by
\begin{equation}
\setlength{\unitlength}{0.007in}%
\begin{picture}(265,189)(100,585)
\thinlines
\put(240,760){\line(-3,-4){120}}
\put(120,600){\line( 1, 0){240}}
\put(360,600){\line(-3, 4){120}}
\put(240,760){\line( 1,-4){ 40}}
\put(260,680){\line(-1, 0){ 80}}
\put(180,680){\line( 1,-4){ 20}}
\put(200,600){\line( 3, 4){ 60}}
\put(260,680){\line( 5,-4){100}}
\put(235,765){\makebox(0,0)[lb]{\raisebox{0pt}[0pt][0pt]{$\alpha_7$}}}
\put(365,595){\makebox(0,0)[lb]{\raisebox{0pt}[0pt][0pt]{$\alpha_8$}}}
\put(100,587){\makebox(0,0)[lb]{\raisebox{0pt}[0pt][0pt]{$\alpha_9$}}}
\put(195,585){\makebox(0,0)[lb]{\raisebox{0pt}[0pt][0pt]{$\alpha_3$}}}
\put(270,585){\makebox(0,0)[lb]{\raisebox{0pt}[0pt][0pt]{$\alpha_2$}}}
\put(155,683){\makebox(0,0)[lb]{\raisebox{0pt}[0pt][0pt]{$\alpha_1$}}}
\put(230,687){\makebox(0,0)[lb]{\raisebox{0pt}[0pt][0pt]{$\alpha_4$}}}
\end{picture}      \label{eq:res1}
\end{equation}
This model is associated with the monomial (\ref{eq:del1}) in the
discriminant.
Denoting the resulting coordinate patch in $\cMc$ by $z_l^{(1)}$ we
obtain
\begin{equation}
\eqalign{z_1^{(1)} &= \frac{a_1a_3a_5a_6}{a_0^3a_9}\cr
z_2^{(1)} &= \frac{a_4a_9}{a_1a_3}\cr
z_3^{(1)} &= \frac{a_3a_7}{a_1a_4}\cr
z_4^{(1)} &= \frac{a_1a_2}{a_3a_4}\cr
z_5^{(1)} &= \frac{a_3a_8}{a_2^2}.\cr}	\label{eq:co1}
\end{equation}
(Note there are no fractional powers of $a_k$ since the large
radius limit point of a smooth \CY\ manifold is a regular point in
$\cMc$, in the example we are considering.)
At this point we can also state the sign in
(\ref{eq:mdmm}) and (\ref{eq:mm}).
We will discuss this issue further in section \ref{ss:per}.
We may associate an integer $d_l$ to each coordinate
$z_l$ defined as the total degree of the numerator or denominator when
expressed in terms of $a_k$. (\ref{eq:mm}) then becomes
\begin{equation}
  B_l + iJ_l = \frac1{2\pi i}\left\{\log\left((-1)^{d_l} z_l
	\right) + O(z_1,\dots,z_5)\right\}  \label{eq:mdmms}
\end{equation}
Thus, in the present example $(-1)^{d_l}=1$ for all $l$.
The $\Box_l$ operators are
\begin{equation}
\eqalign{
  \Box_1 &=\frac\partial{\partial a_1} + \frac\partial{\partial a_3} +
\frac\partial{\partial a_5} +\frac\partial{\partial a_6} -
\frac{\partial^3}{\partial a_0^3} -\frac\partial{\partial a_9}\cr
  \Box_2 &= \frac\partial{\partial a_4} +\frac\partial{\partial a_9}
-\frac\partial{\partial a_1} -\frac\partial{\partial a_3}\cr
  \Box_3 &= \frac\partial{\partial a_3} +\frac\partial{\partial a_7}
-\frac\partial{\partial a_1} -\frac\partial{\partial a_4}\cr
  \Box_4 &= \frac\partial{\partial a_1} +\frac\partial{\partial a_2}
-\frac\partial{\partial a_3} -\frac\partial{\partial a_4}\cr
  \Box_5 &= \frac\partial{\partial a_3} +\frac\partial{\partial a_8}
-\frac{\partial^2}{\partial a_2^2}.\cr}
\end{equation}
We can now write down the differential equations we require
$\Box_l\Phi=0$ by using (\ref{eq:genZ}) and (\ref{eq:co1}).
Rather than attack this daunting set of equations head on we will turn
our attention to sets of ordinary differential equations contained in
this set.

\subsection{Rational curves in $\protect\cMc$.} \label{ss:rc}

The points we are particularly interested in, in this paper, are the
100 points in $\cMc$ which each are the limit of some geometric model,
whether it be smooth \CY, orbifold, Landau-Ginzburg orbifold, etc. As
mentioned in section \ref{ss:cp} toric geometry tells us that the
codimension one walls between the big cones correspond to rational
curves within $\cMc$. In fact, each such rational curve contains
precisely two of our 100 limit points --- the two points given by the
big cones which this wall separates. In our example, each big cone in
$U$ has 5 codimension one faces, that is, given one of the 100 limit
points, there are 5 rational curves in $\cMc$ passing through this
point each of which passes through another limit point. In this way,
there are 250
rational curves which form a ``web'' in $\cMc$ connecting all of the 100
limit points. This is shown as a polytope in figure \ref{fig:web} where lines
represent the rational curves and vertices represent limit points.
(Actually this is the {\em secondary polytope\/} \cite{GZK:sp}). Thus
it is easily seen that one may move in $\cMc$ from any of the limit
points to another one by moving along these rational curves.

\iffigs
\begin{figure}
  \centerline{\epsfxsize=13cm\epsfbox{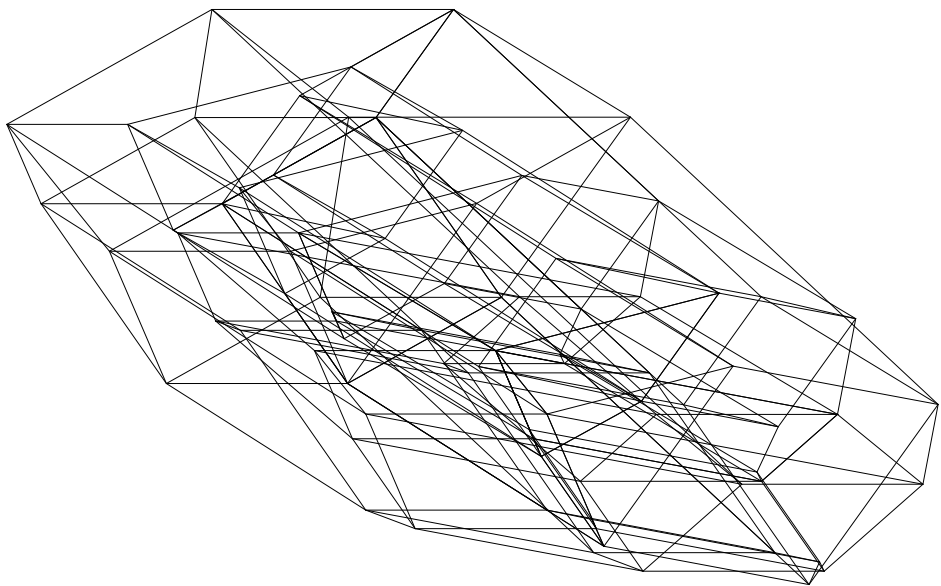}}
  \caption{The web formed by rational curves connecting limit points.}
  \label{fig:web}
\end{figure}
\fi

As we will now see, the form of the set of partial differential
equations from the previous section becomes particularly
straight-forward when restricted to
 these rational curves. We will illustrate this by
an example. Let us consider one of the walls of the cone considered in
(\ref{eq:res1}). A neighbouring cone to this one corresponds to
``resolution 4'' of \cite{AGM:I}, i.e., another smooth \CY\ manifold
this time given by the following triangulation:
\begin{equation}
\setlength{\unitlength}{0.007in}%
\begin{picture}(265,189)(100,585)
\thinlines
\put(240,760){\line(-3,-4){120}}
\put(120,600){\line( 1, 0){240}}
\put(360,600){\line(-3, 4){120}}
\put(240,760){\line( 1,-4){ 40}}
\put(260,680){\line( 5,-4){100}}
\put(260,680){\line(-1, 0){ 80}}
\put(180,680){\line( 5,-4){100}}
\put(180,680){\line( 1,-4){ 20}}
\put(235,765){\makebox(0,0)[lb]{\raisebox{0pt}[0pt][0pt]{$\alpha_7$}}}
\put(365,595){\makebox(0,0)[lb]{\raisebox{0pt}[0pt][0pt]{$\alpha_8$}}}
\put(100,587){\makebox(0,0)[lb]{\raisebox{0pt}[0pt][0pt]{$\alpha_9$}}}
\put(195,585){\makebox(0,0)[lb]{\raisebox{0pt}[0pt][0pt]{$\alpha_3$}}}
\put(270,585){\makebox(0,0)[lb]{\raisebox{0pt}[0pt][0pt]{$\alpha_2$}}}
\put(155,683){\makebox(0,0)[lb]{\raisebox{0pt}[0pt][0pt]{$\alpha_1$}}}
\put(230,687){\makebox(0,0)[lb]{\raisebox{0pt}[0pt][0pt]{$\alpha_4$}}}
\end{picture}      \label{eq:res4}
\end{equation}
For this big cone we have the following coordinates in $\cMc$:
\begin{equation}
\eqalign{z_1^{(2)} &= \frac{a_1a_3a_5a_6}{a_0^3a_9}\cr
z_2^{(2)} &= \frac{a_2a_9}{a_3^2}\cr
z_3^{(2)} &= \frac{a_2a_7}{a_4^2}\cr
z_4^{(2)} &= \frac{a_3a_4}{a_1a_2}\cr
z_5^{(2)} &= \frac{a_1a_8}{a_2a_4}.\cr}
\end{equation}
That is,
\begin{equation}
\eqalign{
 z_1^{(2)} &= z_1^{(1)}\cr
 z_2^{(2)} &= z_2^{(1)}z_4^{(1)}\cr
 z_3^{(2)} &= z_3^{(1)}z_4^{(1)}\cr
 z_4^{(2)} &= \left(z_4^{(1)}\right)^{-1}\cr
 z_5^{(2)} &= z_5^{(1)}z_4^{(1)}.\cr}  \label{eq:tfs}
\end{equation}
The transition functions between these two patches given in
(\ref{eq:tfs}) give us the coordinates for the rational curve
connecting these limit points, i.e., put
$z_1^{(1)}=z_2^{(1)}=z_3^{(1)}=z_5^{(1)}=0$ and
use $z=z_4^{(1)}$ as the coordinate
on the rational curve.

Now let us try to solve $\Box_l\Phi=0$ on this rational curve $C$.
We are interested in finding the regular solution, and the solution
with a $\log(\pm z_4)$-type growth, since their ratio gives the \sm\
coordinate which {\it does not vanish}\/ on $C$.  To eliminate the
other solutions from consideration, we impose the additional
equations
\begin{equation}
  \left.z_n^{(1)} \frac{\partial\Phi}
	{\partial z_n^{(1)}}\right|_C=0, \quad n=1,2,3,5.
     \label{eq:neweqs}
\end{equation}
The solutions to $\Box_l\Phi=0$ with a $\log(\pm z_n)$-type growth for
$n\ne4$, and those that involve powers or products of log terms,
 will fail to satisfy one of these new equations; thus, we will be
left with just the solutions we want.

Using (\ref{eq:neweqs})
immediately allows us to expand out $\Box_4$ in terms of $z$
alone:
\begin{equation}
  \Box_4\Phi = \frac1{a_1a_2}\left\{
	z\frac\partial{\partial z}z\frac\partial{\partial z}
	-z\left(z\frac\partial{\partial z}z\frac\partial{\partial z}
	\right)\right\}\Phi.  \label{eq:de1}
\end{equation}
Now if we consider $\Phi$ as a function of $z$ alone then we have
reduced the problem to an ordinary differential equation.

\subsection{Perestro\u\i ka}	\label{ss:per}

Before trying to solve the differential equation (\ref{eq:de1}) we
will try to generalize the method we followed in the last section so
that we can write down the differential equation for any of the
rational curves in $\cMc$ joining two limit points. To do this we will
first look at the difference between the triangulations of $\cA$
corresponding to the two limit points.

For any $N$, consider $N+2$ points in $\R^N$ such that these points
are not contained in an $\R^{N-1}$ hyperplane.
Let the polytope $Q$ be the convex hull of these points (i.e., the
polytope of minimal volume containing all the points).
It follows
\cite{AxGZ:} that there are precisely two triangulations of this set
of points which contain at least the vertices of $Q$. The transition
between two such triangulations is called a {\em perestro\u\i ka\/}
\cite{GZK:d}. We will give several examples of perestro\u\i ka in later
sections.
The usefulness of the notion of a perestro\u\i ka is that two
triangulations of $\cA$ corresponding to neighbouring cones in our fan
differ by a perestro\u\i ka. That is, we can associate a perestro\u\i ka to
each of the rational curves in $\cMc$ we are considering.

Denoting the $N+2$ points by $\alpha_s$, $s=1,\ldots,N+2$, there will be
a single linear relation between these points
\begin{equation}
  \sum_s m_s \vec\alpha_s = 0,  \label{eq:linrel}
\end{equation}
where $\vec\alpha_s$ is the position vector of $\alpha_s$ and the $m_s$'s
are relatively prime integers. From this relation we define a variable
\begin{equation}
  z = \prod_s a_s^{m_s},
\end{equation}
and a differential operator
\begin{equation}
  \Box = \prod_{m_s>0}\left(\frac\partial{\partial
	a_s}\right)^{m_s} - \prod_{m_s<0}\left(\frac\partial{\partial
	a_s}\right)^{-m_s}.
\end{equation}
Setting
\begin{equation}
  \Phi(a_0,a_1,\ldots) = a_0^{-1}f(z), \label{eq:Phf}
\end{equation}
one can now write $\Box\Phi=0$ as an ordinary differential equation
with $z$ as the only dependent variable.

We claim this construction generalizes that of the previous section.
That is, for any rational curve joining two limit points in $\cMc$ we
can obtain an ordinary differential equation for the periods on $X$ in
terms of $z$, the coordinate on the rational curve. Note that this
ordinary differential equation will always be of {\em hypergeometric
\/} type.

The hypergeometric ordinary differential equations in question on
$\P^1$ has solutions with possible singularities or branch points
at three points which
we will call $z=0,1,\infty$. The points 0 and $\infty$ are the two limit
points which the rational curve connects in $\cMc$. $z=1$ is the only
other singular point and is thus where the
discriminant locus of section \ref{ss:cp} cuts this curve. To be more precise
it is usually the case that the whole rational curve is contained in the
discriminant locus. In this case $z=1$ is the point where another
irreducible component of the discriminant locus cuts\footnote{As we
will see, it is often the case that the a component touches the curve
tangentially rather than cutting transversely or that more that one
component may pass through $z=1$.}
this curve.
Given this form of distinguished points on this curve we can now
specify our choice of branch cuts to perform any analytic
continuation. Each of the points $z=0$ and $z=\infty$ are taken to
represent some limit point around which correlation functions may be
expanded in some power series. This power series fails when one
reaches $z=1$. We thus cut from $z=0$ to $z=1$ and from $z=\infty$ to
$z=1$ to reflect this structure.
(Any other choice would be artificially unsymmetric.)
We extend these cuts from the rational curves on the boundary into
the interior of the moduli space, to form a fundamental domain.
This choice of fundamental domain is implicit in all that
follows in this paper.

In order to put the singularities at $z=1$ we will need to rescale
the $z_l$'s introduced earlier. The sign of this rescaling is the source
of the $(-1)^{d_l}$ factors in the monomial-divisor mirror map. We
may think of this sign as arising from attempting to fix
 the mirror map so that the
number of lines on a \CY\ manifold is positive. As mentioned earlier,
a three-point function in our conformal field theory may be expanded as
an instanton sum in $q_l$
where the coefficients in this series give information
regarding the numbers of holomorphically embedded $\P^1$'s
 on $X$. In particular the
sub-leading term is expected to be
the number of lines (that is, the  number of holomorphically
embedded $\P^1$'s in $X$ the homology class of
whose image is some fixed integral generator
of $H_2(X,\Z)$). To be more precise, in some cases one
may have families of lines depending on parameters,
and then the  ``number of lines'' must be interpreted by means
of the top Chern class of the parameter space of the family
\cite{Witten:tcc,AM:}.  It is generally believed \cite{Gromov:,McD:} that
in such a case a deformation of complex structure to a generic almost
complex structure will yield a discrete set of lines. However, some of
these lines may count negatively\footnote{We thank E.~Witten for
pointing out such a possibility to us.} and thus we cannot
use this strategy to fix
the sign of $z$ in the monomial-divisor mirror map in all cases.
Instead, we first note that if there were
a three-point function whose expansion in $q_l$ had all coefficients
positive, then any pole at the edge of convergence of
such a series would occur when $q_l$ is real and positive.
Since
$q_l=\pm z_l$ to leading order, we find the sign required in such a case
once we know
how to rescale $z_l$ to give a pole at $z=1$. This is the sign choice
$(-1)^{d_l}$ that we specified earlier. By looking at perestro\u\i ka such
that one limit point corresponds to a large radius limit smooth \CY\
manifold, one can show that this sign choice is consistent with the
signs in the principal discriminant given by \cite{GZK:d}.
Unfortunately for a general three-point function, not all of the
coefficients in the $q_l$-expansion need be positive. To maintain
consistency with \cite{GZK:d} we thus {\em conjecture\/} that the
sign given by $(-1)^{d_l}$ is always the correct choice even when negative
coefficients in the expansion occur. That is, we assume that the pole
in the $q_l$-expansion of any 3-point function occurs for a real and
positive $q_l$ (i.e., $B_l=0$).
If our conjecture is wrong and we were to pick the
wrong sign for $z_l$ then we would be counting the number of lines on $X$
with the wrong sign.

In summary we thus do the following. Given the definition of $z_l$ in
(\ref{eq:z2a}) we find the constant by which we need to rescale
$z_l\to z$ to put a pole at $z=1$.
The sign of this factor is $(-1)^{d_l}$.
This sign is absorbed in
the monomial-divisor mirror map so that we only take the absolute
value of this scale factor in our definition of $z$.
Now we will apply this construction to several examples.

It is important to note that although we are describing the
following examples from the perspective of our five-parameter
example, in each case we only actually study the part of the
toric fan specific to the transformation we look at. Thus the
following results are clearly valid for any \CY\ moduli space that is
studied this way. In fact, in string theory, we expect results
concerning flops, blowing-up orbifolds etc., to be dependent only
on the local geometry of $X$. This means that the following examples
should {\em not\/} be considered dependent on the global structure of $X$.

\subsection{The flop}

Recall that a flop is the transformation of a manifold into a
(possibly) topologically different manifold which replaces a $\P^1$ with
another $\P^1$. This occurs by blowing down a $\P^1$ in the original
manifold to form a singular space with a {\em double point}. This
double point can then be resolved by blowing up to give a $\P^1$ in
two different ways. One way returns the original manifold and the
other way yields another manifold. In general a flop need not take a
K\"ahler manifold to another K\"ahler manifold. In this paper however
we are moving from one manifold to another directly by a change of
K\"ahler form and so in this context we are guaranteed a K\"ahler flop.
Any manifold which was a non-K\"ahler flop of $X$ would not have a big
cone in the secondary fan.

The following perestro\u\i ka is a {\em flop}.
\begin{equation}
\setlength{\unitlength}{0.005in}%
\begin{picture}(370,110)(135,675)
\thinlines
\put(140,780){\circle*{10}}
\put(140,680){\circle*{10}}
\put(240,680){\circle*{10}}
\put(240,780){\circle*{10}}
\put(400,780){\circle*{10}}
\put(400,680){\circle*{10}}
\put(500,680){\circle*{10}}
\put(500,780){\circle*{10}}
\put(140,780){\line( 0,-1){100}}
\put(140,680){\line( 1, 0){100}}
\put(240,680){\line( 0, 1){100}}
\put(240,780){\line(-1, 0){100}}
\put(400,780){\line( 0,-1){100}}
\put(400,680){\line( 1, 0){100}}
\put(500,680){\line( 0, 1){100}}
\put(500,780){\line(-1, 0){100}}
\put(140,780){\line( 1,-1){100}}
\put(400,680){\line( 1, 1){100}}
\put(280,730){\vector(-1, 0){  0}}
\put(280,730){\vector( 1, 0){ 80}}
\end{picture}
\end{equation}
This was precisely the perestro\u\i ka considered in section \ref{ss:rc}.
That is, we may specify it by the linear relation (using the numbering
conventions of our example)
\begin{equation}
  \vec\alpha_3+\vec\alpha_4-\vec\alpha_1-\vec\alpha_2=0.
\end{equation}
The ODE associated the the flop, as we saw (in this case no rescaling
of the $z$ parameter is required), is
\begin{equation}
  \left(z\frac{d}{dz}\right)^2f - z\left(z\frac{d}{dz}\right)^2f=0.
\end{equation}
This has a general global solution
\begin{equation}
  f = C_1 + C_2\log(z),
\end{equation}
which is also a general local solution for each $z\neq1$.
We can now follow \cite{Mor:PF} in finding the \smm\ in
terms of $z$. The component of the \smm\ we find is,
of course, the part that varies as we move along the rational curve in
$\cMc$. To find this K\"ahler form, we take the solution for $f(z)$
that behaves to leading order like $\log(z)$ at $z=0$ and divide it
by the solution which is $2\pi i$ to leading order at $z=0$.
That is, we find $B+iJ$ as the ratio of two solutions such that equation
(\ref{eq:mm}) is obeyed.
In this case
this is a trivial task since we have solutions which are exactly a constant and
exactly $\log(z)$. Therefore
\begin{equation}
  B+iJ = \frac1{2\pi i}\log(z).
\end{equation}
That is, the \smm\ is the {\bf same} as the \mKf. To be more
precise {\it
when one performs a flop of one \CY\ manifold into another
and holds all the other components of the K\"ahler form at large radius limit
then the \smm\ coincides with the \mKf}. In particular,
this implies that the area of the flopped $\P^1$ does attain
the value zero in the \sm\ definition, just as it does in the algebraic
definition. In this setting, therefore, string theory does not supply us
with a nonzero lower bound. Of course, the size of the whole manifold
is being kept infinite (i.e., any Riemann surface in a class other
than the one being flopped has infinite area)
and it is only a part of the space which shrinks
to zero.

The flop is a bit too trivial to show the full singularity structure in
the differential equation. One will find however that many three-point
functions will have a pole at $B+iJ=0$.

The fact that the \mKf\ and the \smm\ coincide in the
region of $\cMc$ considered in this section has some interesting
consequences for
the 3-point functions of the superconformal field theory (which we
developed in discussions with Witten \cite{W:phase}). It is known
\cite{OP:flop} that in classical geometry, the K\"ahler cones of two
manifolds related by a flop fit together in $\R^{h^{1,1}}$ by touching
each other along the wall of each K\"ahler cone (where the area of the
flopped $\P^1$ becomes zero). This is equivalent to saying that, so far
as K\"ahler form data is concerned, the class represented by the
flopped $\P^1$ has negative
area in the flopped manifold. Since the \mKf\ generates the same cone
structure as the classical K\"ahler form, the same considerations must
also work for the \mKf\ and thus also for the \smm.

It is important to bear in mind that the homology class of the $\P^1$
present after the flop is the {\it negative\/} of the homology class
present before the flop.  Thus, although the class of the original
$\P^1$ acquires a negative area as the wall between cones is traversed,
the post-flop $\P^1$ will have a positive area in the new region (since
it belongs to the opposite class).

In the portion of $\cMc$ we consider, all Riemann surfaces in $X$,
except for the ones being flopped, are of infinite area. Call the
finite set of $\P^1$'s
being flopped $C_\beta$.
(These are all in the same homology class.)
A 3-point function is then given by
\begin{equation}
  \langle \phi_1\phi_2\phi_3\rangle=
  (D_1\cap D_2\cap D_3) + \sum_\beta \frac{q}{1-q}(D_1\cap C_\beta)
  (D_2\cap C_\beta)(D_3\cap C_\beta), \label{eq:iflp}
\end{equation}
where $q=\exp\{2\pi i(B+iJ)\}$ and $D_n$ is a divisor representing
the field $\phi_n$ in the usual way.

Let us consider the \CY\ manifold $X_1$ at large radius limit. In this
limit $q\to0$ and so the sum in (\ref{eq:iflp}) vanishes. Let us now
flop $X_1$ along the  $C_\beta$'s to obtain the large radius \CY\
manifold $X_2$. Given the discussion above, this is equivalent to
sending $J\to-\infty$, i.e., $q\to\infty$. We can take the proper
transform of the divisors $D_n$ in $X_1$ to obtain divisors in $X_2$
which we denote by the same symbol.
The fundamental equation which relates the intersection numbers before
and after the flop is:
\begin{equation}
  (D_1\cap D_2\cap D_3)_2 = (D_1\cap D_2\cap D_3)_1 - \sum_\beta
  (D_1\cap C_\beta)_1\,(D_2\cap C_\beta)_1\,(D_3\cap C_\beta)_1,
		\label{eq:i12}
\end{equation}
where the subscripts denote in which manifold the intersection numbers
are calculated.
(Note that $C_\beta$ is on $X_1$; we will denote the post-flop $\P^1$'s
by $C_\beta'$.)  Equation
(\ref{eq:i12}) is a statement in classical geometry
which is straightforward to verify.  For example, if we assume that
$D_1$ meets $C_\beta$ transversely at $k$ points, while $D_2$
 and $D_3$ contain $C_\beta$ with multiplicities $l$ and $m$, then
$(D_1\cap D_2\cap D_3)_1=klm$ while $(D_1\cap C_\beta)_1=k$,
$(D_2\cap C_\beta)_1=-l$, and $(D_3\cap C_\beta)_1=-m$.  On the other
hand, after flopping (see figure \ref{fig:flop}), $D_2$ and $D_3$ are
disjoint (at least locally near $C_\beta'$)
so that $(D_1\cap D_2\cap D_3)_2=0$, verifying
(\ref{eq:i12}) in this case.

\begin{figure}

\setlength{\unitlength}{0.01in}%
$$\begin{picture}(349,180)(75,690)
\thinlines
\put( 80,800){\line( 1,-1){ 40}}
\put(120,760){\line(-1,-1){ 40}}
\put(120,760){\line( 1, 0){ 40}}
\put(160,760){\line( 1, 1){ 40}}
\put(160,760){\line( 1,-1){ 40}}
\put(320,700){\line( 1, 1){ 40}}
\put(360,740){\line( 1,-1){ 40}}
\put(360,740){\line( 0, 1){ 40}}
\put(360,780){\line(-1, 1){ 40}}
\put(360,780){\line( 1, 1){ 40}}
\put(350,800){\makebox(0,0)[lb]{\raisebox{0pt}[0pt][0pt]{$D_2$}}}
\put(120,735){\makebox(0,0)[lb]{\raisebox{0pt}[0pt][0pt]{$D_3$}}}
\put(400,755){\makebox(0,0)[lb]{\raisebox{0pt}[0pt][0pt]{$C_\beta^\prime$}}}
\put(140,695){\makebox(0,0)[lb]{\raisebox{0pt}[0pt][0pt]{$C_\beta$}}}
\put(395,760){\vector(-1, 0){ 30}}
\put(150,715){\vector( 0, 1){ 40}}
\multiput(210,760)(7.82609,0.00000){12}{\line( 1, 0){  3.913}}
\put(300,760){\vector( 1, 0){0}}
\put(210,760){\vector(-1, 0){0}}
\put( 75,755){\makebox(0,0)[lb]{\raisebox{0pt}[0pt][0pt]{$D_1$}}}
\put(125,775){\makebox(0,0)[lb]{\raisebox{0pt}[0pt][0pt]{$D_2$}}}
\put(350,705){\makebox(0,0)[lb]{\raisebox{0pt}[0pt][0pt]{$D_3$}}}
\put(325,755){\makebox(0,0)[lb]{\raisebox{0pt}[0pt][0pt]{$D_1$}}}
\put(240,765){\makebox(0,0)[lb]{\raisebox{0pt}[0pt][0pt]{flop}}}
\put(125,850){\makebox(0,0)[lb]{\raisebox{0pt}[0pt][0pt]{$X_1$}}}
\put(350,850){\makebox(0,0)[lb]{\raisebox{0pt}[0pt][0pt]{$X_2$}}}
\end{picture}$$

  \caption{A flop.}
  \label{fig:flop}
\end{figure}

If we now calculate the 3-point function (\ref{eq:iflp}) using (\ref{eq:i12}),
we find
\begin{equation}
\eqalign{
        \langle \phi_1\phi_2\phi_3\rangle_1&=
        (D_1\cap D_2\cap D_3)_2 + \sum_\beta \left(\frac{q}{1-q}+1\right)
        (D_1\cap C_\beta)_1(D_2\cap C_\beta)_1(D_3\cap C_\beta)_1 \cr
        &=(D_1\cap D_2\cap D_3)_2 + \sum_\beta \frac{q^{-1}}{q^{-1}-1}
        (-D_1\cap C_\beta')_2(-D_2\cap C_\beta')_2(-D_3\cap C_\beta')_2 ,
        }
         \label{eq:instcalc}
\end{equation}
where we have used $(D\cap C_\beta)_1=(-D\cap C_\beta')_2$.  Noting that the
change in sign of homology class $[C_\beta']=-[C_\beta]$ demands that
we replace $q$ by $q^{-1}$, we conclude that
$\langle \phi_1\phi_2\phi_3\rangle_1=\langle \phi_1\phi_2\phi_3\rangle_2$,
as expected.

\subsection{The $\hbox{\bigbbbfont Z}_2$-quotient singularity}
	\label{ss:Z2}

The following is the only perestro\u\i ka in one-dimension:
\begin{equation}
\setlength{\unitlength}{0.005in}%
\begin{picture}(310,150)(195,655)
\thinlines
\put(200,800){\circle*{10}}
\put(200,660){\circle*{10}}
\put(500,660){\circle*{10}}
\put(500,800){\circle*{10}}
\put(200,730){\circle*{10}}
\put(500,730){\circle{10}}
\put(200,800){\line( 0,-1){140}}
\put(500,800){\line( 0,-1){140}}
\put(280,730){\vector(-1, 0){  0}}
\put(280,730){\vector( 1, 0){140}}
\end{picture}  \label{eq:Z2}
\end{equation}
In this picture the network on the left has 2 lines whereas on the
right the middle point is ignored and there is only one line. In our
example we have a few such configurations, e.g.,
\begin{equation}
  \vec\alpha_3+\vec\alpha_8-2\vec\alpha_2=0.
\end{equation}
Indeed this perestro\u\i ka can be applied to (\ref{eq:res1}) by removing
the point $\alpha_2$ from the triangulation. The model of $X$ thus
obtained has a curve of $\Z_2$ quotient singularities \cite{AGM:II}.
The operation (\ref{eq:Z2}) is (moving from right to left)
precisely the resolution of a $\Z_2$ quotient singularity in $\C^2$
where the $\Z_2$ action in $\C^2$ is $(z_1,z_2)\mapsto(-z_1,-z_2)$.

The rational curve associated to the perestro\u\i ka (\ref{eq:Z2}) thus
joins a limit point of a space with a $\Z_2$ quotient singularity to
the limit point of a space where such a singularity has been blown-up
(to infinite size).

To put a branch-point at $z=1$ we define
\begin{equation}
  z = 4\frac{a_3a_8}{a_2^2},	\label{eq:thisZ2}
\end{equation}
i.e., we have introduced a factor of 4. The associated ODE is
\begin{equation}
  \left(z\frac{d}{dz}\right)^2f - z\left(z\frac{d}{dz}\right)
	\left(z\frac{d}{dz}+\ff12\right)f=0.
\end{equation}
This has a general solution
\begin{equation}
  f=C_1 + C_2\log\left(\frac{2-z-2\sqrt{1-z}}z\right).
\end{equation}
Clearly the term which is constant at $z=0$ is again exactly constant.
Expanding the second term around $z=0$ we obtain (assuming the square
root to be positive)
\begin{equation}\label{eq:theabove}
  \log\left(\frac{2-z-2\sqrt{1-z}}z\right) =
\log(z/4)+{\ff {1}{2}}z+{\ff {3}{16}}z^{2}+{\ff {5}
{48}}z^{3}+{\ff {35}{512}}z^{4}+{\ff {63}{1280}}z^{5}+O\left (z^{6
}\right).
\end{equation}
Because of our rescaling of the $z$ variable we now need to look for a
solution which behaves like $\log(z/4)$ to leading order. This is
simply given by (\ref{eq:theabove}). Thus we obtain
\begin{equation}
  B+iJ=\frac1{2\pi i}\log\left(\frac{2-z-2\sqrt{1-z}}z\right).
\end{equation}

This therefore gives an example where the \smm\ and the
\mKf\ do {\em not\/} agree. An interesting question we can ask is what
is the value of $B+iJ$ at the orbifold limit point, i.e., when
$z\to\infty$. The component of the K\"ahler form we are studying is
the class controlling the areas of Riemann surfaces which lie
 in the exceptional divisor resulting
from blowing-up this singularity. That means that in some sense we
are looking at the
volume of this exceptional divisor. Na\"\i vely of course from
classical geometry we assume that this volume is zero at the orbifold
point but we see that the \mKf\ would have us believe
that the relevant areas are $-\infty$. To find what the \smm\
tells us let us introduce the variable
\begin{equation}
  \psi=z^{-1/2}
\end{equation}
and then carefully rewrite $B+iJ$ in this variable assuming $0<\arg(\psi)
<\pi$ to obtain
\begin{equation}
  B +iJ=-\frac1\pi\cos^{-1}\psi,
\end{equation}
where we take the branch corresponding to $\cos^{-1}\psi=
\frac\pi2-\psi+O(\psi^3)$.  (This is consistent with our earlier
choice of branch cuts.)
Thus we see that the orbifold, given by $\psi=0$ corresponds to
$B=-1/2$ and $J=0$. The fact that $J=0$ means that the \smm\
agrees with the classical volume, i.e., the volume of the
exceptional divisor (and thus the areas of the Riemann surfaces within it)
before you blow-up a singularity is zero. More
curious is the value $B=-1/2$ at the orbifold point which would appear
to have no classical explanation.

Note that we have measured the volume of the exceptional divisor at
the the limit point of the orbifold in $\cMc$, that is, all sizes
except that associated with the exceptional divisor are infinite. It
is an interesting question to see whether the exceptional divisor has
non-zero volume in the case of an orbifold {\em not\/} at an
otherwise large radius limit. We hope to address this question in
future work.\footnote{Recently a two parameter moduli space has been
studied in detail \cite{CDFKM:I} which should help address this question.}

Proponents of a universal ``$R\to1/R$'' symmetry in the moduli space
of string vacua should take note that in passing from the smooth
blown-up \CY\ manifold to the orbifold we have been able to shrink the
Riemann surfaces within the
exceptional divisor completely down to zero size without being able to
identify this with some equivalent large radius model. There is no
symmetry between the orbifold points and any other
points in the moduli space. Thus it
would appear that string theory does {\em not\/} remove all distances less than
the Planck scale from a moduli space. Some parts of a target space can
become as small as they wish at least so long as the rest of the
target space is at large radius limit.

In this example in moving from the
\mKf\ to the \smm\ we have removed negative
areas. That is, $J\geq0$ for all points on the rational curve in
$\cMc$. We will discuss this further after looking at some more
orbifolds.

The branch-point of the general solution of this hypergeometric equation
is at $z=\psi=1$, i.e., $B+iJ=0$. This is where many three-point functions
will diverge in the conformal field theory. This shows that the only
difference between an orbifold, where string theory is known to be
well behaved, and a ``bad'' conformal theory is the value of the
$B$-field since in both cases $J=0$, i.e., the volume of the exceptional
divisor is zero.

Let us now look at the form of the discriminant for this $\Z_2$
resolution in the context of our example. The monomial for the
large-radius limit resolution is given by (\ref{eq:del1}) and one may
derive the monomial in $\Delta_p$ corresponding to the neighbouring
cone in the secondary fan corresponding to the \CY\ space with
a curve of $\Z_2$-quotient singularities as
\begin{equation}
  r_\xi\delta_\xi = -64a_0^{18}a_1^{8}a_3^{11}a_4^{10}
	a_5^{12}a_6^{12}a_7^{6}a_8^{9}a_9^{4}. \label{eq:del1-2}
\end{equation}
If we assert that the discriminant locus intersects our rational curve
in $\cMc$ at $z=1$ then we see immediately that
$\Delta_p$ for points in $\cMc$ near this rational curve is given by
\begin{equation}
\eqalign{\Delta_p&\simeq a_0^{18}a_1^{8}a_2^{6}a_3^{8}a_4^{10}
	a_5^{12}a_6^{12}a_7^{6}a_8^{6}a_9^{4}\,(1-z)^3\cr
  &= a_0^{18}a_1^{8}a_2^{6}a_3^{8}a_4^{10}
	a_5^{12}a_6^{12}a_7^{6}a_8^{6}a_9^{4} -
     12a_0^{18}a_1^{8}a_2^{4}a_3^{9}a_4^{10}
	a_5^{12}a_6^{12}a_7^{6}a_8^{7}a_9^{4} \cr
     &\qquad\qquad+
     48a_0^{18}a_1^{8}a_2^{2}a_3^{10}a_4^{10}
	a_5^{12}a_6^{12}a_7^{6}a_8^{8}a_9^{4} -
     64a_0^{18}a_1^{8}a_3^{11}a_4^{10}
	a_5^{12}a_6^{12}a_7^{6}a_8^{9}a_9^{4}.\cr}
\end{equation}
This shows how important terms from $\widetilde\Delta_p$ are on the
rational curves in $\cMc$. In this case we derive two terms in
$\widetilde\Delta_p$, i.e., terms in $\Delta_p$ which could not
be obtained by the methods of section \ref{ss:cp}.

In this paper we will usually use the word ``orbifold'' to refer to a
space whose only singularities are locally of the form of quotient
singularities. It is more conventional when talking about conformal
field theories to consider an orbifold to be {\em globally\/} of the
form of a quotient of a smooth manifold (or conformal field theory).
In this case one can determine the massless spectrum of the theory to be
composed of fields from the original smooth theory combined with
twisted fields from the quotient singularities in the new space.
(Massive fields can also appear from group elements with no fixed
points.) The specific example of a curve of $\Z_2$-quotient singularities
 we have
considered for $z$ given by (\ref{eq:thisZ2}) cannot be globally written
as an orbifold. Despite this fact we claim that we can still relate
more conventional conformal field theory ideas to this orbifold as we
now argue.

Consider the Landau-Ginzburg orbifold theory given by the minimal triangulation
of $\cA$ given by the simplex
$\alpha_5\alpha_6\alpha_7\alpha_8\alpha_9$.
This is an orbifold theory in the conformal field theory sense and
thus has a ``quantum''-symmetry group \cite{Vafa:qs} isomorphic to the
group by which we quotiented the original Landau-Ginzburg model.
This
$\Z_{18}$ symmetry is given by $x_4\to\exp(2\pi i/18)x_4$. The monomial
$a_1x_2^3x_4^9$ transforms as a faithful representation of a $\Z_2$
subgroup of this group. Thus if this monomial is added to the
Landau-Ginzburg superpotential we would break the $\Z_2$ symmetry.
This is precisely the conformal field theory picture of resolving a
$\Z_2$-quotient singularity --- we add the twisted marginal operator
$a_1x_2^3x_4^9$ into the action to break the discrete symmetry.
In terms of toric geometry this resolution of a singularity in $X$
corresponds to a subdivision of the fan representing $X$ by
a star subdivision (see for example \cite{AGM:II}). Such a subdivision
adds a point in $\cA$ into the triangulation. By the monomial-divisor
mirror map, this point in $\cA$ is
precisely the point that represents the monomial which acts as the
twisted marginal operator --- i.e., $\alpha_1$.

In our example we do not have a global quotient singularity but it is
locally of the form of a quotient singularity and thus we expect at
least the massless part of the conformal field theory to behave as if
it were an orbifold. This is because massless twist fields can be
considered to be localized around the fixed points.
Thus we claim that for the transition given by
(\ref{eq:thisZ2}) the ``twisted'' marginal operator is the monomial
corresponding to the point in $\cA$ added into the triangulation by
 the perestro\u\i ka --- namely $a_2x_3^6x_4^6$.
Indeed if we follow the approach of \cite{AGM:II} to find which
superpotential, i.e., which values of $a_k$, give
the relevant space with a $\Z_2$-quotient
singularity we find this consistent with $a_2=0$, i.e., this
marginal operator switched off.

Take the theory with a quotient singularity and perform
a perturbative expansion for small values of $a_2$ to blow-up the
singularity (much along the lines of \cite{Cve:orb} for example).
Our 3-point functions will be in the form of
a power series in $a_2$, i.e., $\psi
=a_2/\sqrt{4a_3a_8}$
if we write things in a $(\C^*)^5$-invariant way. We know that the
discriminant locus occurs at $\psi=1$ and so this marks the boundary
of the circle of convergence for such a power series. In particular
such a perturbative method cannot reach the smooth target space
($\psi\to\infty$) before breaking down.

This is one way of viewing the ``phases'' picture of the moduli space
\cite{W:phase}. In one region of moduli space containing the orbifold
theory we may use perturbation theory in twisted (perhaps only in the
local sense) marginal operators to calculate all 3-point functions.
This region is neighboured by another region containing the point
corresponding to the quotient singularity having been resolved with an
exceptional divisor of infinite size. Any 3-point function in this
region may be calculated by an expansion in terms of instantons given
by $\P^1$'s in the exceptional divisor. On the boundary between
these two regions the twisted marginal field's coefficient becomes too
large for the twisted field perturbation theory to converge and on
the other hand the exceptional divisor becomes too small for the
instanton expansion on $\P^1$'s to converge.

\subsection{The $\hbox{\bigbbbfont Z}_3$-quotient singularity}
	\label{ss:Z3}

Consider the following perestro\u\i ka in $\R^2$:
\begin{equation}
\setlength{\unitlength}{0.005in}%
\begin{picture}(525,130)(85,665)
\thinlines
\put(170,790){\circle*{10}}
\put(250,670){\circle*{10}}
\put( 90,670){\circle*{10}}
\put(170,710){\circle*{10}}
\put(170,790){\line( 2,-3){ 80}}
\put(250,670){\line(-1, 0){160}}
\put(170,790){\line(-2,-3){ 80}}
\put( 90,670){\line( 2, 1){ 80}}
\put(170,710){\line( 0, 1){ 80}}
\put(170,710){\line( 2,-1){ 80}}
\put(525,790){\circle*{10}}
\put(605,670){\circle*{10}}
\put(445,670){\circle*{10}}
\put(525,710){\circle{10}}
\put(280,730){\vector(-1, 0){  0}}
\put(280,730){\vector( 1, 0){140}}
\put(525,790){\line( 2,-3){ 80}}
\put(605,670){\line(-1, 0){160}}
\put(525,790){\line(-2,-3){ 80}}
\end{picture}
\end{equation}

A fan based on the these triangles gives the toric description of an isolated
$\Z_3$-quotient singularity and its blow-up. The quotient singularity in $\C^3$
is given by the action $(z_1,z_2,z_3)\mapsto(\omega z_1,\omega
z_2,\omega z_3)$ where $\omega=\exp(2\pi i/3)$. In our example this
perestro\u\i ka can occur based on the following relation
\begin{equation}
   \vec\alpha_1+\vec\alpha_7+\vec\alpha_8-3\vec\alpha_4=0.
\end{equation}
One of the smooth models of $X$ (resolution number 5 in \cite{AGM:I})
admits this perestro\u\i ka and so one of the big cones neighbouring this
cone corresponds to a target space that has acquired a $\Z_3$ quotient
singularity.

Defining
\begin{equation}
  z = -27\frac{a_1a_7a_8}{a_4^3},
\end{equation}
we obtain the differential equation
\begin{equation}
  \left(z\frac{d}{dz}\right)^3f - z\left(z\frac{d}{dz}\right)
   \left(z\frac{d}{dz}+\ff13\right)\left(z\frac{d}{dz}+\ff23\right)
	f=0. \label{eq:Z3h}
\end{equation}
Again the solution that is regular at $z=0$ is just a constant. This
time however the solution that behaves like $\log(z)$ cannot be
determined in terms of elementary functions.

To find the required solution of (\ref{eq:Z3h}) we need to turn to the
theory of hypergeometric functions. Indeed, with the exception of the
flop and the $\Z_2$-orbifold, all the ODE's we obtain for a
perestro\u\i ka will require hypergeometric function theory to find the
\smm.

Recall \cite{Slater:hyp} that the hypergeometric function $\F{N+1}N$ is
defined by the infinite series
\begin{equation}
\eqalign{
  \F{N+1}N(a_1,a_2,\ldots,a_{N+1};\,b_1,b_2,&\ldots,b_N;\,z)=
  1+\frac{a_1a_2\ldots a_{N+1}}{b_1b_2\ldots b_{N}} \frac{z}{1!}\cr
  &+\frac{a_1(a_1+1)a_2(a_2+1)\ldots a_{N+1}(a_{N+1}+1)}
  {b_1(b_1+1)b_2(b_2+1)\ldots b_{N}(b_{N}+1)}\frac{z^2}{2!}+\ldots,\cr
}
\end{equation}
where $a_n,b_n$ are complex numbers (not to be confused with any
previous use of these symbols). This series converges for $|z|<1$.
The following ODE has as a solution
$f(z)=\F{N+1}N(a_1,\ldots;\,b_1,\ldots;\,z)$:
\begin{equation}
\eqalign{
  \Biggl\{z\frac{d}{dz}\left(z\frac{d}{dz}+b_1-1\right)&
  \left(z\frac{d}{dz}+b_2-1\right)\ldots
  \left(z\frac{d}{dz}+b_N-1\right)\cr
  &-z\left(z\frac{d}{dz}+a_1\right)\left(z\frac{d}{dz}+a_2\right)
  \ldots\left(z\frac{d}{dz}+a_{N+1}\right)
  \Biggr\}f=0.\cr}  \label{eq:hDE}
\end{equation}

All the differential equations encountered when finding the \smm\
are of the form (\ref{eq:hDE}). Thus applying
hypergeometric theory to our differential equation (\ref{eq:Z3h}) we
obtain the solution $f(z)=\F32(0,\ff13,\ff23;\,1,1;\,z)=1$. Hence we
recover the solution we already knew.

To find the other solution we require, we substitute
\begin{equation}
  g(z) = z\frac{d}{dz}f(z)
\end{equation}
into (\ref{eq:Z3h}). This leads to a lower-order hypergeometric
differential equation with solution $g(z)=\F21(\ff13,\ff23;\,1;\,z)$.
Expanding this solution we obtain
\begin{equation}
\eqalign{f(z) &= \int z^{-1}g(z)\,dz\cr
	&= \int z^{-1}\left(1+\ff29z+\ff{10}{81}z^2+
		\ff{560}{6561}z^3+\ldots\right)\,dz\cr
	&= \log(z) + C + \ff29z + \ff5{81}z^2 +
		\ff{560}{19683}z^3+\ldots,\cr} \label{eq:fint}
\end{equation}
for some constant, $C$. This clearly provides the other solution we
require so that
\begin{equation}
  B+iJ = \frac1{2\pi i}\left\{\log(z/27)+ \ff29z + \ff5{81}z^2 +
	\ff{560}{19683}z^3+\ldots\right\}  \label{eq:Z3s}
\end{equation}

Now let us determine, as we did for the $\Z_2$ quotient singularity,
the areas of the Riemann surfaces within
the exceptional divisor when $z\to\infty$. In the case
of the $\Z_2$ orbifold we have an exact form for the \smm\
and so this determination was straight-forward. In the $\Z_3$
case however we only have a series solution and this clearly diverges
as $z\to\infty$. What we require therefore is the analytic
continuation of the series in (\ref{eq:Z3s}). This may be done by
solving the hypergeometric differential equation, this time as a
series expanded around $z=\infty$ and then match these solutions to
the solutions around $z=0$. This is known as the {\em connection
problem} (see for example \cite{IKSY:}).
In \cite{CDGP:} the connection problem was solved by finding
a complete set of periods as a series solution around $z=0,1,\infty$
and then demanding that the transformation between these be
symplectic. We shall employ another method which is more
straight-forward to apply to a general case.

The connection problem is simple to solve with Barne's-type integrals
when no solutions with
logarithmic poles are involved. Consider our function $g(z)$ above
represented as a Barne's-type integral:
\begin{equation}
  \F21(\ff13,\ff23;\,1;\,z)=\frac1{2\pi i\Gamma(\ff13)\Gamma(\ff23)}
  \int_{-i\infty}^{+i\infty}\frac{\Gamma(t+\ff13)\Gamma(t+\ff23)
  \Gamma(-t)}{\Gamma(t+1)}(-z)^t\,dt,
\end{equation}
where $|\arg(-z)|<\pi$. The integration path moves to the left around
the pole at $t=0$ as shown in figure \ref{fig:path}.
\begin{figure}

\setlength{\unitlength}{1mm}
$$\begin{picture}(100,80)(0,0)
\thinlines
\put(0,40){\line(1,0){100}}
\put(50,0){\line(0,1){80}}

\multiput(48,38)(10,0){6}{\makebox(4,4){$\times$}}
\multiput(44.66666,38)(-10,0){5}{\makebox(4,4){$\times$}}
\multiput(41.33333,38)(-10,0){5}{\makebox(4,4){$\times$}}

\thicklines
\put(50,0){\line(0,1){38.3}}
\put(50,80){\line(0,-1){38.3}}
\put(50,40){\oval(3.2,3.2)[l]}
\put(50,10){\vector(0,1){10}}
\put(50,50){\vector(0,1){10}}

\put(15,70){\makebox(0,0){$t$-plane}}

\end{picture}$$

  \caption{The integration path for $\F21(\ff13,\ff23;\,1;\,z)$.}
  \label{fig:path}
\end{figure}
The series form of this hypergeometric function is recovered if one
completes the integration path into a loop to enclose all the poles to
the right of the path. The residues at the non-negative integers form
the infinite sum. To find these residues and also to prove many of the
following relations in this paper we use
\begin{equation}
  \Gamma(x)\Gamma(1-x) = \frac\pi{\sin(\pi x)}. \label{eq:Gid}
\end{equation}

One may also complete the path to the left enclosing the poles at
$n-\ff13,n-\ff23$, where $n$ is a non-positive integer. This expresses
our hypergeometric function as another sum which now converges for
$|z|>1$. This new sum is therefore the analytic continuation of the
original sum.
In fact this new sum is a sum of other hypergeometric
functions:
\begin{equation}
\eqalign{
\F21(\ff13,\ff23;\,1;\,z) \simeq \frac{\Gamma(\ff13)}{\Gamma^2(\ff23)}
  e^{-\frac{\pi i}3}\psi\,&\F21(\ff13,\ff13;\,\ff23;\,\psi^3)\cr
  &-3\frac{\Gamma(\ff23)}{\Gamma^2(\ff13)}
  e^{-\frac{2\pi i}3}\psi^2\,\F21(\ff23,\ff23;\,\ff43;\,\psi^3),\cr}
	\label{eq:F12c}
\end{equation}
where ``$\simeq$'' denotes analytic continuation and we have
introduced $\psi=z^{-1/3}$ such that $0<\arg(\psi)<2\pi/3$. For
details see, for example, page 136 of \cite{Slater:hyp}.

To analytically continue our definition of $B+iJ$ to the orbifold
point we need to multiply (\ref{eq:F12c}) by $z^{-1}$ and integrate.
Doing this directly would introduce an integration constant which
would be undetermined. To answer the question of what the size of
the exceptional divisor at the orbifold point is, we need to know this
constant. Let us instead na\"\i vely apply this process directly to
the integrand in the Barne's-type integral. This gives the following function:
\begin{equation}
  h(z) = \frac1{2\pi i\Gamma(\ff13)\Gamma(\ff23)}
  \int_{-i\infty}^{+i\infty}\frac{\Gamma(t+\ff13)\Gamma(t+\ff23)
  \Gamma(-t)}{t\Gamma(t+1)}(-z)^t\,dt. \label{eq:Bl1}
\end{equation}
Completing this path to the right and writing it as a sum of residues
we certainly recover the part of (\ref{eq:fint}) which is a power
series in $z$. The subtlety arises because of the double pole we now
have at $t=0$. Remember that if $w(t)$ is nonzero and finite at $t=0$
then the residue of $w(t)/t^2$ at $t=0$ is $(w^\prime(t))_{t=0}$. It
follows that
\begin{equation}
  h(z) = \log(-z) + \Psi(\ff13) + \Psi(\ff23) - 2\Psi(1)
	+\ff29z + \ff5{81}z^2 + \ff{560}{19683}z^3+\ldots,
\end{equation}
where $\Psi(n)$ is the {\em digamma\/} or {\em psi\/} function
which is defined as the derivative of $\log\Gamma(n)$. Since
(from 8.365.6 of \cite{GR:big})
\begin{equation}
\sum_{k=1}^{n-1}\Psi(k/n)-(n-1)\Psi(1)
	= -n\log n,
\end{equation}
we have
\begin{equation}
  h(z) = \log(-\frac{z}{27})+\ff29z + \ff5{81}z^2 +
	\ff{560}{19683}z^3+\ldots 	\label{eq:hasp}
\end{equation}
Thus
\begin{equation}
  B+iJ=\frac1{2\pi i}h(z) -\ff12.
\end{equation}
We can now analytically continue $B+iJ$ into the $|z|>1$ region by
completing the path of the integral in (\ref{eq:Bl1}) to the left and
writing it as a sum over residues. The result is
\begin{equation}
\eqalign{
  h(z) = -3\frac{\Gamma(\ff13)}{\Gamma^2(\ff23)}
  e^{-\frac{\pi i}3}\psi\,&\F32(\ff13,\ff13,\ff13;\,\ff23,\ff43;\,\psi^3)\cr
  &+\frac92\frac{\Gamma(\ff23)}{\Gamma^2(\ff13)}
  e^{-\frac{2\pi i}3}\psi^2\,\F32(\ff23,\ff23,\ff23;\,
	\ff43,\ff53;\,\psi^3).\cr} \label{eq:hpsi}
\end{equation}
Note this is a combination of hypergeometric functions which
are solutions to third-order differential equations. These equations
may be derived directly from (\ref{eq:Z3h}) by suitable changes of
variable.

Putting $\psi=0$ to obtain the orbifold point we see that $B+iJ=-\ff12$.
Thus the size of the exceptional divisor is again zero and again the
$B$-field has value $-\ff12$.
Notice the cancellation that was required between the digamma
functions appearing from the double pole in the Barne's-type integral
and the $\Vol(\sigma)^{\Vol(\sigma)}$-type factors that were required
to achieve this seemingly trivial final result.

Note from (\ref{eq:hpsi}) that, for $|\psi|\ll1$ we have
$\pi/6<\arg(\ff1{2\pi i}h(z))<5\pi/6$. This shows that $J\geq0$, i.e., we
have only non-negative areas in this region. In fact, negative areas
are completely excluded from the conformal field theories
parameterized by this rational curve in $\cMc$.

We have also done enough to determine the value of $B+iJ$ at $z=1$ where
we expect the conformal field theory to be singular. One finds $B=0$ and
$J\approx0.463$ ($\approx 18.3\alpha^\prime$ putting back units of
length). Thus, in contrast to the $\Z_2$-singularity case, the
discriminant now vanishes when we acquire a specific non-zero size for
the exceptional divisor.

Naturally everything we said about the description of a theory in
terms of twisted marginal operators in the last section also applies to
this case. In our example the twisted marginal operator resolving the
$\Z_3$-quotient singularity would be $a_4x_2^3x_3^3x_4^3$. Again this
is only locally of the form of a quotient singularity and this
operator is not twisted under any global symmetry of a covering
theory.

The moduli space of K\"ahler forms on the so-called $Z$-manifold was
studied in \cite{Drk:Z}. This manifold is the resolution of an
orbifold with $\Z_3$-quotient singularities of the form studied in
this section. Indeed similar hypergeometric functions appear in
\cite{Drk:Z} where the entire region of moduli space in the
orbifold ``phase'' is studied.

\subsection{The $\hbox{\bigbbbfont Z}_4$-quotient singularity}

In this paper we will concentrate mainly on perestro\u\i ka which take one
from a large radius limit smooth \CY\ manifold to a neighbouring cone.
This is because we know how to define the K\"ahler form for the smooth
\CY\ manifold by using the monomial-divisor mirror map. If we look at any
other perestro\u\i ka it would be necessary to first determine $B+iJ$ at
one of the limit points by following a path from a smooth \CY\
manifold limit point. As we discuss briefly later, such a path will
usually raise considerations about basis changes as one moves from one
perestro\u\i ka to the next. In this section we look at a simple example
where we may ignore such problems. That is, we will blow down two
irreducible divisors which will not ``interfere'' with each other and
thus no basis change is required.

Any quotient singularity in \CY\ spaces of complex dimension 3
other than the two we have just studied will require an exceptional
divisor with more than one irreducible component. This means that a
complete resolution of the singularity requires more than one
perestro\u\i ka.
Consider the next simplest case:
\begin{equation}
\setlength{\unitlength}{0.005in}%
\begin{picture}(490,340)(80,460)
\thinlines
\put(165,795){\circle*{10}}
\put(245,675){\circle*{10}}
\put(165,675){\circle*{10}}
\put(165,735){\circle*{10}}
\put( 85,675){\circle*{10}}
\put(485,795){\circle*{10}}
\put(565,675){\circle*{10}}
\put(485,675){\circle*{10}}
\put(405,675){\circle*{10}}
\put(485,585){\circle*{10}}
\put(565,465){\circle*{10}}
\put(405,465){\circle*{10}}
\put(165,585){\circle*{10}}
\put(245,465){\circle*{10}}
\put(165,525){\circle*{10}}
\put( 85,465){\circle*{10}}
\put(485,525){\circle{10}}
\put(485,465){\circle{10}}
\put(165,465){\circle{10}}
\put(485,735){\circle{10}}
\put(260,740){\vector(-1, 0){  0}}
\put(260,740){\vector( 1, 0){120}}
\put(165,655){\vector( 0, 1){  0}}
\put(165,655){\vector( 0,-1){ 50}}
\put(260,525){\vector(-1, 0){  0}}
\put(260,525){\vector( 1, 0){120}}
\put(485,655){\vector( 0, 1){  0}}
\put(485,655){\vector( 0,-1){ 50}}
\put(165,795){\line( 0,-1){120}}
\put(165,795){\line( 2,-3){ 80}}
\put(245,675){\line(-1, 0){160}}
\put( 85,675){\line( 2, 3){ 80}}
\put( 85,675){\line( 4, 3){ 80}}
\put(165,735){\line( 4,-3){ 80}}
\put(485,795){\line( 0,-1){120}}
\put(485,795){\line( 2,-3){ 80}}
\put(565,675){\line(-1, 0){160}}
\put(405,675){\line( 2, 3){ 80}}
\put(485,585){\line( 2,-3){ 80}}
\put(565,465){\line(-1, 0){160}}
\put(405,465){\line( 2, 3){ 80}}
\put(165,585){\line( 2,-3){ 80}}
\put(245,465){\line(-1, 0){160}}
\put( 85,465){\line( 2, 3){ 80}}
\put( 85,465){\line( 4, 3){ 80}}
\put(165,525){\line( 4,-3){ 80}}
\put(165,585){\line( 0,-1){ 60}}
\put(465,620){\makebox(0,0)[lb]{\raisebox{0pt}[0pt][0pt]{1}}}
\put(320,745){\makebox(0,0)[lb]{\raisebox{0pt}[0pt][0pt]{2}}}
\put(320,530){\makebox(0,0)[lb]{\raisebox{0pt}[0pt][0pt]{3}}}
\put(140,625){\makebox(0,0)[lb]{\raisebox{0pt}[0pt][0pt]{4}}}
\end{picture}   \label{eq:Z4t}
\end{equation}
The bottom-right diagram is the toric picture for a $\Z_4$-quotient
singularity in $\C^3$ generated by $(z_1,z_2,z_3)\mapsto
(iz_1,iz_2,-z_3)$. The top-left diagram is the complete blow-up of
this singularity to give a smooth space. There are two irreducible
components to the exceptional divisor and thus two perestro\u\i ka are
involved. The two components of the exceptional divisor may be
produced in either order in the blow-up procedure so that there are
two possible paths the perform the blow-up as shown above.

Such a choice of paths is a common feature in $\cMc$. It is clear from
figure \ref{fig:web} that the journey between any two limit points may
be taken along many paths. In order for us to be able to give a value
of the \smm\ to each limit point we require that the
choice of paths does not affect this value.

One of the paths in this $\Z_4$ example (taken along line 2 and 1 in
(\ref{eq:Z4t})) consists of two perestro\u\i ka of the type considered in
section \ref{ss:Z2}. We know therefore that this path leads to zero
volumes for both components of the exceptional divisor in the orbifold
limit. In the alternative path, line 4 is also of this type so that
one component of the exceptional divisor is again zero at the orbifold
point. In order for (\ref{eq:Z4t}) to be commutative we thus require
that the perestro\u\i ka given by line 3 gives zero volume at this point
as we will now check.

In our example this configuration is given by
\begin{equation}
   \vec\alpha_3+2\vec\alpha_7+\vec\alpha_8-4\vec\alpha_4=0,
\end{equation}
leading to
\begin{equation}
  z=64\frac{a_3a_7^2a_8}{a_4^4}.
\end{equation}
We can now follow the procedure in section \ref{ss:Z3} where now we
are dealing with the solutions of the equation related to the
hypergeometric function $\F43(0,\ff14,\ff12,\ff34;\,\ff12,1,1;\,z)$.
Again the regular solution is a constant and again we obtain the other
solution by the trick in equation (\ref{eq:Bl1}). This time we obtain
\begin{equation}
  \eqalign{h(z)&=\log(-z/64) +\ff3{16}z +\ff{105}{2048}z^2+
	\ff{385}{16384}z^3+\ldots\cr
  &=-4\frac{\Gamma(\ff14)}{\Gamma^2(\ff34)}e^{-\frac{\pi i}4}
	\psi\,\F43(\ff14,\ff14,\ff14,\ff34;\,
	\ff12,\ff34,\ff54;\,\psi^4)+\ldots,\cr}
\end{equation}
where $\psi=z^{-1/4}$ and $0<\arg(\psi)<\pi/2$. Thus at the orbifold
point, as $\psi\to0$ we have zero volume again as we expected.

\subsection{Changing the dimension of $X$}

Thus far we have always obtained the the result that $B+iJ=-\ff12$
at the limit point where we remove a point from $\cA$ from the
triangulation. This is {\em not\/} a general feature. It is not
difficult to convince oneself however that it will be when can use
the constructions above. That is, the digamma functions introduced by
the double pole at $z=0$ will cancel the factor introduced in the
definition of $z$. This construction of the \smm\ relied
on the fact that one of the solutions of the hypergeometric
differential equation was a constant.

Remember from (\ref{eq:Phf}) that $a_0$ plays a distinguished r\^ole
in our hypergeometric system. So far, non of the perestro\u\i ka
considered have involved the point $\alpha_0$. So long as this is
true, we will obtain a hypergeometric equation with a constant
solution. In fact, it is not hard to prove that the condition for
{\em not\/} having a
constant solution is as follows. When the linear relation
(\ref{eq:linrel}) is formed, $m_0$ must be non-zero and have opposite
sign to the other non-zero $m_s$'s. This statement is equivalent to
the statement that the associated perestro\u\i ka consists of removing
(or adding) $\alpha_0$ to the triangulation.

It was shown in \cite{AGM:II} that the point $\alpha_0$ plays a
distinguished r\^ole for another reason. If $\alpha_0$ is a vertex of
every simplex in the triangulation of $\cA$ then $X$ may be
interpreted as an irreducible space of complex dimension 3. If
$\alpha_0$ is a vertex of only some of the simplices then $X$ is
reducible with only part of $X$ having a 3-dimensional representation. If
$\alpha_0$ does not appear in the triangulation then the dimension of
$X$ is $<3$. Thus, the perestro\u\i ka we have not yet considered are the
ones which lower the dimension of $X$ down from 3.

An extreme example of this is effectively the one studied in
\cite{CDGP:} where the other limit point is a Landau-Ginzburg orbifold theory,
i.e., $X$ has dimension 0. We shall first study one of the neighbours
of ``resolution 5'' of \cite{AGM:I} where the dimension is shrunk
down to 2 (see \cite{AGM:II} for a full explanation of this). The
perestro\u\i ka is identical to that considered in section \ref{ss:Z3}
except now the relation is
\begin{equation}
   \vec\alpha_4+\vec\alpha_5+\vec\alpha_6-3\vec\alpha_0=0,
\end{equation}
and thus
\begin{equation}
  z = -27\frac{a_4a_5a_6}{a_0^3}.
\end{equation}
Now the differential equation is
\begin{equation}
  \eqalign{
  \left(z\frac{d}{dz}\right)^3f - &z\left(z\frac{d}{dz}+\ff13\right)
   \left(z\frac{d}{dz}+\ff23\right)\left(z\frac{d}{dz}+1\right)f\cr
  &=\left(z\frac{d}{dz}\right)\left\{\left(z\frac{d}{dz}\right)^2f
  -z\left(z\frac{d}{dz}+\ff13\right)\left(z\frac{d}{dz}+\ff23\right)f
  \right\}\cr
  &=0.\cr}        \label{eq:Z3h0}
\end{equation}
Thus, the solution which is regular at $z=0$ is given by $f(z)=
\F21(\ff13,\ff23;\,1;\,z)$. In (\ref{eq:F12c}) we gave the analytic
continuation of this for $|z|>1$. We now need to find the solution of
(\ref{eq:Z3h0}) that behaves as $\log(z)$ at $z=0$ and continue this
to $|z|>1$.

{}From our earlier analysis of Barne's-type integrals we saw that a
double pole gave a residue with a $\log(z)$ term. With this success in
mind consider the following
\begin{equation}
  h(z) = -\frac1{2\pi i\Gamma(\ff13)\Gamma(\ff23)}
  \int_{-i\infty}^{+i\infty}\Gamma(t+\ff13)\Gamma(t+\ff23)
  \Gamma^2(-t)z^t\,dt. \label{eq:Bl2}
\end{equation}
Completing the path to the right and taking residues we obtain
\begin{equation}
\eqalign{
  h(z)=\F21(\ff13,\ff23;\,1;\,z)&\log(z) - \log(27)\cr
	&+\frac1{\Gamma(\ff13)\Gamma(\ff23)}\sum_{n=1}^\infty
	\left[\frac\partial{\partial t}\left(\frac
	{\Gamma(t+\ff13)\Gamma(t+\ff23)}{\Gamma^2(t+1)}\right)
	\right]_{t=n}z^n.\cr}
\end{equation}
Completing the path to the left and taking residues we obtain
\begin{equation}
\eqalign{
  h(z)=-\frac{\Gamma(\ff13)}{\Gamma^2(\ff23)}\frac\pi
	{\sin(\frac\pi3)}&\psi\,\F21(\ff13,\ff13;\,\ff23;\,\psi^3)\cr
	&+3\frac{\Gamma(\ff23)}{\Gamma^2(\ff13)}\frac\pi
	{\sin(\frac\pi3)}\psi^2\,\F21(\ff23,\ff23;\,\ff43;\,\psi^3),\cr}
		\label{eq:F12nc}
\end{equation}
where, as in section \ref{ss:Z3}, $\psi=z^{-1/3}$ and $0<\arg(\psi)
<2\pi/3$.
The above expresses $h(z)$ as a linear combination of the same
functions that appeared in (\ref{eq:F12c}) and so $f(z)=h(z)$ is a
solution of (\ref{eq:Z3h0}). Thus we have found the solution that
behaves like $\log(z)$ at $z=0$ and its analytic continuation for
$|z|>1$. This is a general method for finding the extra solutions of a
hypergeometric equation whose regular solution is
$\F{N+1}N(a_1,a_2,\ldots;\,1,1,\ldots;\,z)$ --- simply take some of
the $\Gamma(t+1)$ factors in the denominator of the Barne's-type
integral and move them into the numerator as $\Gamma(-t)$ terms (with
a change in sign of $z$ for each term). This produces a high-order
pole at $z=0$ which gives some power of $\log(z)$ when the residue is
taken.

The monomial-divisor mirror map tells us
\begin{equation}
  B+iJ=\frac1{2\pi i}\frac{h(z)}{\F21(\ff13,\ff23;\,1;\,z)}.
\end{equation}
To find the value of $B+iJ$ at the limit point corresponding to the
2-dimensional target space, we take $|\psi|\ll1$ whence from
(\ref{eq:F12c}) and (\ref{eq:F12nc}) we obtain
\begin{equation}
  B+iJ\simeq\frac{ie^{\frac{\pi i}3}}{2\sin\frac\pi3}\left\{
	1-3\frac{\Gamma^3(\ff23)}{\Gamma^3(\ff13)}(1-
	e^{-\frac{\pi i}3})\psi+O(\psi^2)\right\}.
\end{equation}
Thus for $\psi=0$ we have $B=-\ff12$ and $J=\ff12\cot(\pi/3)$. That
is, the area of the generator of $H_2(X,\Z)$ (and thus we infer the
volume of $X$)
at this limit point is {\em not\/} zero. We also see
from the above expression that for small $\psi$ we have
\begin{equation}
\pi/6<\arg\left((B+iJ)-(B+iJ)_{\psi=0}\right)<5\pi/6,
\end{equation}
showing how the size always increases as we move away from $\psi=0$.

The method of calculation we have just done may also be applied to the
mirror of the quintic threefold as studied in \cite{CDGP:}. In this
case we obtain the result that at the Landau-Ginzburg orbifold point we obtain
$J=\ff12\cot(\pi/5)$ (which is equal to $\frac45\sin^3(2\pi/5)$ as
stated in \cite{CDGP:}).

We thus see that it is impossible to shrink the whole \CY\ manifold
down to a point as measured by the \smm. If we think of
a path on figure \ref{fig:web} that begins at a smooth \CY\ point and
ends on the Landau-Ginzburg orbifold point then one of the lines we traverse
must correspond to a perestro\u\i ka that removes $\alpha_0$ and hence
yields $J>0$. Notice that to calculate the value of $B+iJ$ at, say,
the Landau-Ginzburg orbifold point is quite complicated. As we follow a path
along the web of figure \ref{fig:web}, at each vertex we have to change the
basis of $B+iJ$ to prepare for the next perestro\u\i ka.
Because, as mentioned, earlier we do not have a nice linear structure
on the moduli space expressed in terms of the \smm\
coordinates, the basis change will generally involve transcendental functions.

Consider the perestro\u\i ka that adds the point $\alpha_0$ to the minimal
triangulation comprising of just the simplex $P^\circ$. This is the
transition between the Landau-Ginzburg orbifold and the \CY\ space which
is a hypersurface in the unresolved $\P^4_{\{6,6,3,2,1\}}$ (see
\cite{AGM:II} for more details). If we measure the volume of the
Landau-Ginzburg orbifold according to this transition by the above
calculation we obtain $J=\ff12\cot(\pi/18)$. This shows that the above
value of $J=\ff12\cot(\pi/3)$ must change as we blow-down the
remaining parts of $X$ to obtain the Landau-Ginzburg orbifold. This
change occurs because of the basis changes in this process.


\section{Discussion and Conclusion}  \label{s:conc}

We began this paper by noting two properties of string theory which
are relevant for understanding the space of allowed target space metrics.
First, recent work \cite{AGM:I,AGM:II,W:phase}\ has shown that
string theory makes sense even if the target space metric does not
satisfy the usual positivity conditions that one classically expects.
In this regard we are led to augment the space of allowed
K\"ahler forms beyond the usual K\"ahler cone.
Second, a number of works  have demonstrated that string theory appears
to impose ``minimal lengths" and hence restricts the physically
relevant space of K\"ahler forms to lie within the usual K\"ahler cone.
Part of the purpose of the present work has been to study these
divergent tendencies and show that, in fact, they are completely
consistent.

In particular, since the concept of ``size" is an intrinsically classical
mathematical notion, we have carefully studied ways of extending
its meaning to the more abstract realm of conformal field theory. In essence,
we have sought to find natural continuations of the definition of
size from classical to quantum geometry. There is no unique way of doing
this. We have found, though, that when our conformal field theory
has a sigma model interpretation we can extract a definition of size,
by using mirror symmetry,
from the geometric structure of the latter.
We can then extend this definition by analytic continuation to
all theories in the enlarged K\"ahler moduli space.
We have seen, in particular, that this gives rise to a precise meaning,
rooted in the structure of conformal sigma models, to the {\it area
of two-cycles\/} throughout the moduli space.

With sufficient calculational power, we would explicitly carry out
this program and thereby study the full realm of possible
areas for these cycles. The work of
\cite{AGM:I,AGM:II,W:phase},
for example, would appear to indicate that zero and negative areas
would necessarily arise. The present paper, though, has shown that
the definition of area that one would directly extract from these
works (the \mKf) does not agree with the natural sigma model
K\"ahler form discussed above. We have therefore sought to
determine if the latter definition restores something akin to the
usual positivity conditions. For calculational ease, we have limited
our attention to particular complex dimension one subspaces in the
enlarged K\"ahler moduli space for the illustrative example studied
in \cite{AGM:I,AGM:II}.

In terms of the \mKf\ the  enlarged moduli space of K\"ahler forms
naturally leads to many ``K\"ahler
cones'', each associated to its own geometric model of $X$, glued
together spanning the whole $\R^{h^{1,1}}$. In our example there are
100 such cones of which 5 correspond to smooth \CY\ manifolds.
We have considered complex dimension one spaces in this moduli
space which join the ``large radius limit points'' in each region.
In order to determine the value of the \smm\ at each of
these limit points, we have considered the network of rational curves in
the compactification divisor of the moduli space which connects them.
 Each such rational curve leads to an ordinary
hypergeometric equation allowing for an analysis along the lines
of \cite{CDGP:}. It is reasonable to expect that the extreme
values of the \smm\ will occur at the these limit points.
In this paper we have demonstrated this only in a limited sense by
looking at the neighbourhoods within rational curves
of some of the limit points, where we discover that {\it not}\/ all values
in $\R^{h^{1,1}}$ are  attained by  the \smm.
To be more precise, all
necessarily negative\footnote{By ``necessarily negative'' areas we mean
areas which are negative when measured with respect to {\it all}\/ of
the birational models $X_i$ of $X$.}
areas are eliminated as well as small positive sizes where all
components of the \smm\ are small. Notice however that
some Riemann surfaces {\em can\/} be shrunk down to zero area while
other parts of $X$ are held at large-radius limit.

More precisely, if we plot the ``$J$'' part of the
\smm\ (i.e., the
imaginary part of $B+iJ$) in $\R^{h^{1,1}}$ we do not find a cone
structure for each of the 100 phase regions.
The 5 cones of the smooth \CY\ models are retained as cones asymptotically away
from the origin since the \smm\ and the \mKf\ coincide
there. All of the other 95 limit points are mapped somewhere within
these 5 cones --- i.e., they all have non-negative areas with respect to at
least one of these 5 models. Thus, assuming these limit points
represent extreme values of $J$, the whole moduli space of \smm s
maps into this union of 5 cones.
We can represent this idea in figure~\ref{fig:mush}. We show roughly how the
space of algebraic $J$'s as shown in figure~\ref{fig:as} is expected
to be modified
in going to the same diagram of the \smm\ $J$ in a hypothetical example
where two of the cones give smooth \CY\ manifolds. The two smooth \CY\
regions are labeled $X_1$ and $X_2$ and the other regions are
labeled a, b and c.

\iffigs
\begin{figure}
  \centerline{\epsfxsize=15cm\epsfbox{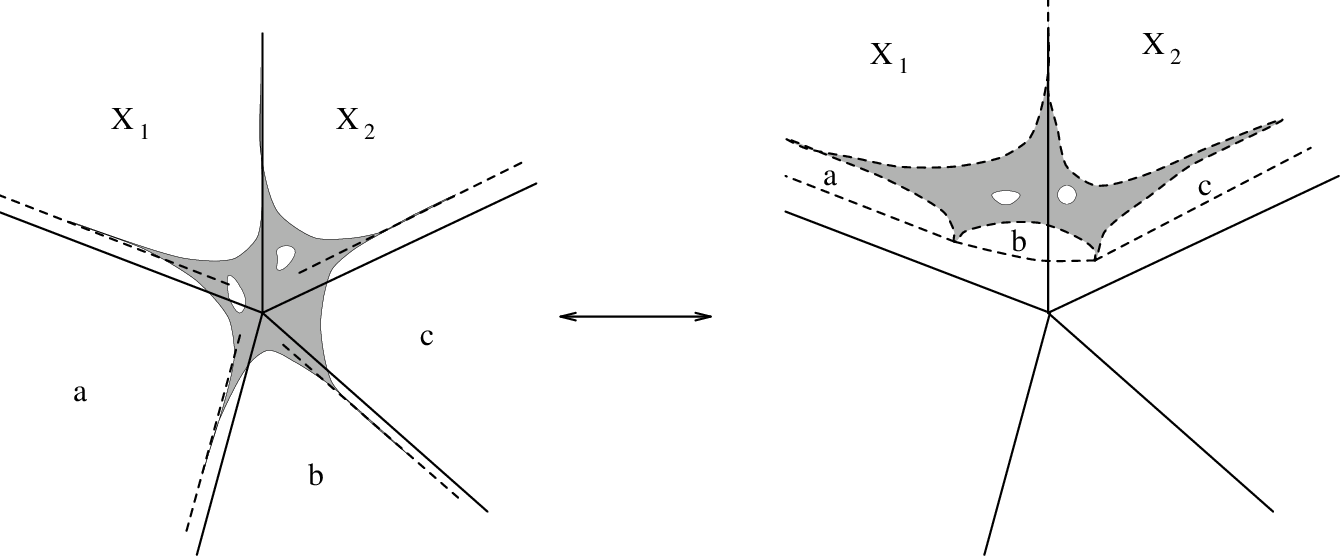}}
  \caption{The different phases as they appear in $J$-space for}
\vskip 2pt
\centerline{\parbox{2.75in}{the \mKf\ (on the left) and
the \smm\ (on the
right).}}
  \label{fig:mush}
\end{figure}
\fi

It would appear therefore that no negative areas appear in the space
of conformal field theories describing non-linear \sm s or at least if
negative areas do occur then they can be redefined away by using
a topologically different model for $X$. Thus, string theory does require
us to enlarge the space of allowed K\"ahler forms beyond the usual
classical K\"ahler cone, but it does so in a manner consistent with
non-negative areas. In this way, we resolve the puzzle discussed at
the beginning of this section and in the introduction.

It is interesting to compare the results of this paper with the results
of classical general relativity --- i.e., the moduli space of Ricci-flat
metrics on $X$. It turns out that that both the flop and the quotient
singularity (and its blow-up) appear as limiting classical solutions
to the Einstein equations. First we
describe the flop.
This case was studied in \cite{CD:flop} and we shall repeat here only
an outline of the argument. With a $2\times2$ matrix representation
of the coordinates, $W$, one can define a distance $r^2=\tr(W^\dagger
W)$ from the double point when the $\P^1$ is blown down. The metric
can then be written
\begin{equation}
  ds^2=\cF\,{}^\prime\tr(dW^\dagger\,dW)+\cF\,{}^{\prime\prime}
\left|\tr(dW^\dagger\,dW)\right|^2+4c\frac{|d\lambda|^2}{1+|\lambda|^2},
\end{equation}
where $\cF$ is some function of $r$ and $\lambda$ is the coordinate on
the $\P^1$.
The real parameter $c$ in the above corresponds to the area of the flopped
$\P^1$ and thus may be taken as the component of the (real) K\"ahler
form which gives this area. If $c=0$ then the above metric is degenerate
(in the sense that some
distinguished points are now separated by zero distance).
If $c<0$ one may change coordinates to give a smooth metric with
a $\P^1$ with area $-c$ \cite{CD:flop}. Thus we see that
the only difference between
this picture and the stringy picture we presented in terms of $\cMc$ is
that we have an extra degree of freedom in the $B$-field which may be
used to smooth out the singularity at $c=0$ as far the conformal field
theory is concerned \cite{AGM:I}.

When we turn to the quotient singularity there is a bigger difference.
For example, let us consider the blow-up of the singularity of the type
considered in section \ref{ss:Z3}.
Near the $\Z_3$ singularity, or its blow-up, we have \cite{FG:}:
\begin{equation}
  ds^2 = 2\frac{(c^3+L^3)^{\frac13}}L\left\{dx^id\bar x^i
  -\frac{c^3}{L(c^3+L^3)}\bar x^idx^i x^jd\bar x^j\right\},
\end{equation}
where $L=x^i\bar x^i$.
This contains a real parameter $c>0$ for a smooth metric (with
suitable change of coordinates).
This is roughly the form of the metric irrespective of the global
geometry so long as $x,\sqrt{c}\ll R$, where $R$ is some characteristic
length of the global geometry of $X$.
As shown in \cite{FG:} this geometry contains a $\P^2$ submanifold
with the standard Fubini-Study metric. The line element, $ds^2$, on this
$\P^2$ is proportional to the parameter $c$. This $\P^2$ is clearly
the exceptional divisor with $c=0$ giving the quotient singularity.
Varying this parameter thus corresponds to varying the component of
the real K\"ahler form that gives the volume of the exceptional
divisor. For a smooth metric we require $c>0$. This gives the
classical moduli space a boundary. If we continue into the $c<0$
region then the $\P^2$ acquires negative size and part of $X$ becomes
``pinched off''. This metric is still Ricci-flat and so depending on
one's qualms about negative areas one might wish to consider this a
solution of classical general relativity.

When we look at the orbifold point in $\cMc$ we see a different picture.
Now the point in $\cMc$ which corresponds to a zero volume exceptional
divisor is not on a boundary --- $\cMc$ has no boundary. As we move in
any direction away from this point the volume of the exceptional
divisor becomes positive (or remains zero) and so we never need to
address the question of negative volumes.
We are unable to pinch off regions of space in this manner.
In general the singularities in the conformal field theories appear in
a different location in the moduli space compared to the classical
picture. In the case of a $\Z_2$-quotient singularity the singular
theory appears at zero-volume exceptional divisor, i.e., just where the
singular metric occurs but for the $\Z_3$-quotient singularity we need
a small but non-zero exceptional divisor to have a singular theory. That is,
the string theory is singular when the classical theory is smooth!

We should be clear about our language of which regions are and are not
included in our moduli space. The situation is analogous to the
string on a circle of radius $R$ and the $R\leftrightarrow1/R$
duality. One point of view is to say that distances $<1$ exist but may
be reinterpreted as distances $>1$. The other point of view is to say
that string theory cuts off distances $<1$. We are implicitly assuming
this second point of view in the above. This is because we have
{\em defined\/} the \smm\ in terms of the large radius
limit(s) of the \sm\ and thus have fixed ourselves in the $R>1$ region.
Any ruler which can measure distances on our large radius circle
manifold and give the correct answer will be unable to measure
distances $<1$.

To regain the $R\leftrightarrow1/R$ picture of any distance existing
but some being equivalent, one would take the moduli space $\cMc$
and form the simply-connected, smooth covering space. The group by
which one mods this covering space out by to form $\cMc$ is the
{\em modular group\/} of the target space $X$. While there is nothing
wrong with such a process from the mathematical point of view one
should ask what the physical meaning of such a construction is. In
terms of the \smm\ we have introduced negative areas
but declared at the same time that they are entirely equivalent to
positive areas. Do such areas really exist? The only way that they
can be measured is to define the method of reading a ruler such that
we get answers which would not agree with the large radius limit \CY\
manifold but this is precisely where we try to make contact with our
classical ideas of distance!

In conclusion we have shown that when building the moduli space of
allowed \smm, all distances, at least for the limit
points, are non-negative. In this statement we mean non-negative when
measured according to at least one of the smooth models of $X$. Some
of the limit points, such as large-radius limit orbifolds, do admit
zero distances however showing that string theory does not cut off
distances shorter than the Planck scale.

\section*{Acknowledgements}

We thank E. Witten for helpful discussions.
P.S.A.\ would like to thank J. Schiff for useful conversations and for
reminding him that one can sometimes solve partial differential equations.
B.R.G.\ would like to thank S. Hosono for calling attention to the
utility of \cite{Bat:var}.
The work of P.S.A.\ was supported by DOE grant
DE-FG02-90ER40542, the work of B.R.G.\ was supported
by a National Young Investigator award, the Ambrose Monell Foundation and
the Alfred P. Sloan Foundation,
and the work of D.R.M.\ was supported
by an American Mathematical Society Centennial Fellowship.

\pagebreak

\end{document}